\documentclass[aps,prx,twocolumn,superscriptaddress,floatfix,showkeys]{revtex4-2}
\usepackage{etoolbox}
\usepackage[english]{babel}
\usepackage{graphics,graphicx} 
\usepackage{natbib}
\usepackage[labelformat=parens]{subfig}
\usepackage{float}
\usepackage{placeins} 
\usepackage{ragged2e}  
\usepackage{caption}
\usepackage{xcolor}
\usepackage{amsmath}
\usepackage{amssymb}
\usepackage{braket}
\usepackage{blindtext}
\usepackage{bm}
\usepackage{algorithm}
\usepackage{algpseudocode}

\usepackage{hyperref}
\hypersetup{
    colorlinks=true,
    citecolor=blue, 
    linkcolor= red, 
    urlcolor= blue 
}
\usepackage[labelformat=parens]{subfig}
\usepackage[justification=raggedright, font=footnotesize]{caption} 

\usepackage{MnSymbol}

\usepackage[cal=boondox,scr=boondoxo]{mathalfa}
\usepackage{listings}

\definecolor{codegreen}{rgb}{0,0.6,0}
\definecolor{codegray}{rgb}{0.5,0.5,0.5}
\definecolor{codepurple}{rgb}{0.58,0,0.82}
\definecolor{backcolour}{rgb}{0.95,0.95,0.92}

\lstdefinestyle{mypython}{
    backgroundcolor=\color{backcolour},   
    commentstyle=\color{codegreen},
    keywordstyle=\color{magenta},
    numberstyle=\tiny\color{codegray},
    stringstyle=\color{codepurple},
    basicstyle=\ttfamily\footnotesize,
    breakatwhitespace=false,         
    breaklines=true,                 
    captionpos=b,                    
    keepspaces=true,                 
    numbers=left,                    
    numbersep=5pt,                  
    showspaces=false,                
    showstringspaces=false,
    showtabs=false,                  
    tabsize=2,
    frame=single,
    framesep=5pt,
    rulecolor=\color{black},
    xleftmargin=10pt,
    xrightmargin=10pt
    basewidth=0.48\textwidth
}

\begin{document}

\title{ Bridging Frustration and Non-Hermiticity via COMPASS: An Adaptive Biorthogonal Neural Quantum State Framework}

\author{Lavoisier Wah}
\email[]{lavoisier.wahkenounouh@mpl.mpg.de}
\affiliation{Max Planck Institute for the Science of Light, 91058 Erlangen, Germany}
\affiliation{Department of Physics, Friedrich-Alexander-Universit\"at Erlangen-N\"urnberg, 91058 Erlangen, Germany}

\author{Flore K. Kunst}
\affiliation{Max Planck Institute for the Science of Light, 91058 Erlangen, Germany}
\affiliation{Department of Physics, Friedrich-Alexander-Universit\"at Erlangen-N\"urnberg, 91058 Erlangen, Germany}

\author{Mohamed Hibat-Allah}
\affiliation{Department of Applied Mathematics, University of Waterloo, Waterloo, ON N2L 3G1, Canada}
\affiliation{Vector Institute, Toronto, Ontario, M5G 0C6, Canada}

\date{\today}

\begin{abstract}
In this work, we introduce a complementary optimization method for progressive and adaptive state search (COMPASS) based on biorthogonal adaptive recurrent neural quantum states. Our approach combines an adaptive autoregressive architecture with a biorthogonal variational Monte Carlo scheme as well as a complementary optimization scheme that alternates adiabatically between energy and variance minimization. This enables the stable convergence to ground-state eigenpairs with high fidelity, while avoiding Markov chain sampling through exact autoregressive generation. We demonstrate that for parity-time($PT$)-symmetric Hamiltonians, unconstrained complex ans\"{a}tze can spontaneously break $PT$ symmetry during optimization--even in the unbroken phase--leading to spurious imaginary energies. Real-valued ans\"{a}tze in an appropriate basis, on the other hand, naturally constrain the optimization to the correct physical manifold. Conversely, for generic non-Hermitian (NH) Hamiltonians without symmetry protection and complex spectra, complex ans\"{a}tze are essential for accurately capturing complex ground-state properties. Our results establish that physically-informed ansatz selection is crucial for reliable NH simulations. By combining adaptive architectures, biorthogonal optimization, and symmetry-aware modeling, this framework enables a direct study of 1D and 2D NH many-body systems without Hermitian embeddings or adiabatic continuation. Applying this framework to systems with frustrated magnetism, we show that gap frustration provides a quantitative shield against NH spectral instability, with the frustration gap setting a critical threshold for $PT$-symmetry breaking. At the same time, complexifying the frustration coupling itself generates a new topologically nontrivial network of diabolic level crossings, controlled by the phase of the complex coupling, that has no Hermitian analog. We then establish the emergence of a novel spectral topology in NH frustrated systems: the diabolic ring.

\end{abstract}

\maketitle

\section{Introduction}\label{S0}

Neural quantum states (NQSs)~\cite{troyer2017} have proven to be a versatile variational ansatz for strongly correlated many-body systems \cite{PhysRevResearch.2.023358,torlai2018neural, PhysRevLett.125.100503, medvidovic2025adiabatic}, offering expressive representations that scale favorably with system size. While early applications focused primarily on Hermitian Hamiltonians~\cite{PhysRevLett.125.100503, medvidovic2024neural, medvidovic2025adiabatic}, NQS methods have recently been extended to non-Hermitian (NH) systems by explicitly representing their many-body wavefunction~\cite{wah2025, solinas2025biorthogonal}. These advances have enabled the study of NH ground states beyond exact diagonalization (ED) \cite{wah2025,solinas2025biorthogonal}, including in higher dimensions and near exceptional points (EPs) \cite{solinas2025biorthogonal}. However, most existing approaches rely on fixed network architectures \cite{wah2025,solinas2025biorthogonal}, Markov chain Monte Carlo sampling \cite{solinas2025biorthogonal}, and unconstrained complex-valued ans\"atze \cite{zhang2025observation, wah2025, solinas2025biorthogonal}, which can introduce optimization instabilities and hidden biases. 

In parallel, adaptive neural quantum states have recently been proposed as a way to dynamically adjust the model capacity during training, enabling faster convergence, improved training stability, and better scaling with system size \cite{mcnaughton2025adaptive, zen2020transfer}. By progressively increasing the expressive power of the network, adaptive architectures act as a form of curriculum learning that mitigate optimization pathologies commonly encountered in large variational spaces. Despite their success in Hermitian settings, the role of adaptive architectures in NH quantum systems and their interplay with biorthogonality and symmetry remain largely unexplored. Extending this approach to an NH setting is the raison d'\^etre of this study.

A key conceptual challenge in NH systems is that the traditional variational Monte Carlo (VMC) is no longer valid because the left and right eigenstates are no longer orthogonal but rather biorthogonal. To circumvent this problem, previous studies proposed a variance-based optimization process that guarantees convergence of the algorithm to the correct pair of biorthogonal eigenstates~\cite{solinas2025biorthogonal, siringo2005variational, PhysRevLett.94.150201, xie2024variational} . Then, to guarantee that these pairs of eigenstates are effectively the ground-state wavefunction of the system, they proposed mapping the NH Hamiltonian to a Hermitian Hamiltonian and adiabatically adding non-Hermiticy \cite{solinas2025biorthogonal}.

In this work, we introduce a \emph{complementary optimization method for progressive adaptive state search (COMPASS)} based on biorthogonal adaptive recurrent neural quantum states. Our approach combines three key ingredients: (i) an adaptive autoregressive recurrent neural network~(RNN) architecture that progressively increases the model capacity during training, (ii)~a biorthogonal NQS representation of the left and right eigenstates, which unlike previous approaches restricted to $PT$-symmetric Hamiltonians, enables the direct treatment of generic NH Hamiltonians, which has not been studied so far; and (iii) a complementary optimization process that alternates between energy minimization and variance minimization to ensure convergence to the correct ground-state eigenpair. Importantly, our autoregressive formulation enables exact sampling without Markov chains, eliminating autocorrelation effects and improving the stability near critical and exceptional points. By combining adaptive architectures with complementary biorthogonal optimization and symmetry-aware modeling, our work provides the first principled framework for directly studying one-dimensional~(1D) and two-dimensional~(2D) NH many-body systems without relying on Hermitian embeddings, adiabatic continuations, or symmetry simplification, thereby granting access to a class of problems that has, until now, remained entirely beyond the reach of existing methods. This positions NQSs as a robust and physically informed tool for exploring the rich landscape of NH systems.

Moreover, while the interplay between non-Hermiticity and quantum many-body physics has emerged as a frontier in condensed matter theory only recently \cite{Ashida02072020}, frustrated quantum magnetism has been a central theme in condensed matter physics for decades~\cite{PhysRevB.43.13559, Lhuillier2001,mathew2025quantum}. The one-dimensional spin-$1/2$ $J_1-J_2$ chain is a paradigmatic model of frustrated magnetism, exhibiting a Berezinskii-Kosterlitz-Thouless quantum phase transition at $J_2/ J_1\approx 0.2411$--the Okamoto-Nomura (ON) critical point~\cite{OKAMOTO1992433}-- from a gapless spin-liquid phase to a gapped dimerized phase, with an exactly solvable Majumdar-Ghosh (MG) point at $J_2/ J_1=0.5$ \cite{mg10}. Despite the extensive separate developments in both fields, the intersection of NH physics and frustrated magnetism remains largely unexplored. Existing work on NH spin chains has focused on integrable and stoquastic models without frustration~\cite{wah2025,solinas2025biorthogonal}, interacting fermionic systems~\cite{bbDMRG5vnlw9p4}, and Yang-Lee edge singularities~\cite{GvonGehlen_1991, PhysRevLett.40.1610}. Crucially, the effect of non-Hermiticity on frustration-driven quantum phases, such as dimerization and the ground state at the MG point, has not been addressed. Here, we fill this gap by introducing two physically motivated NH extensions of the $J_1-J_2$ chain, thereby applying our COMPASS framework to systems beyond stoquastic, non-frustrated models and opening a new avenue for exploring the interplay between frustration and non-Hermitian physics.

This paper is organized as follows. In Sec.~\ref{S1}, we present our framework and the 1D models on which the experiments are performed. In Sec.~\ref{S2}, we demonstrate that the choice of the ansatz is crucial in the context of NH systems. Then, we extended the study to 2D models in Sec.~\ref{S3}, and to frustrated models in Sec.~\ref{S4}. Finally, our work is summarized in Sec.~\ref{S5}.

\section{ Method and Comparisons}\label{S1}

\subsection{Method}

\begin{figure*}
    \centering
    \includegraphics[width=0.99\textwidth]{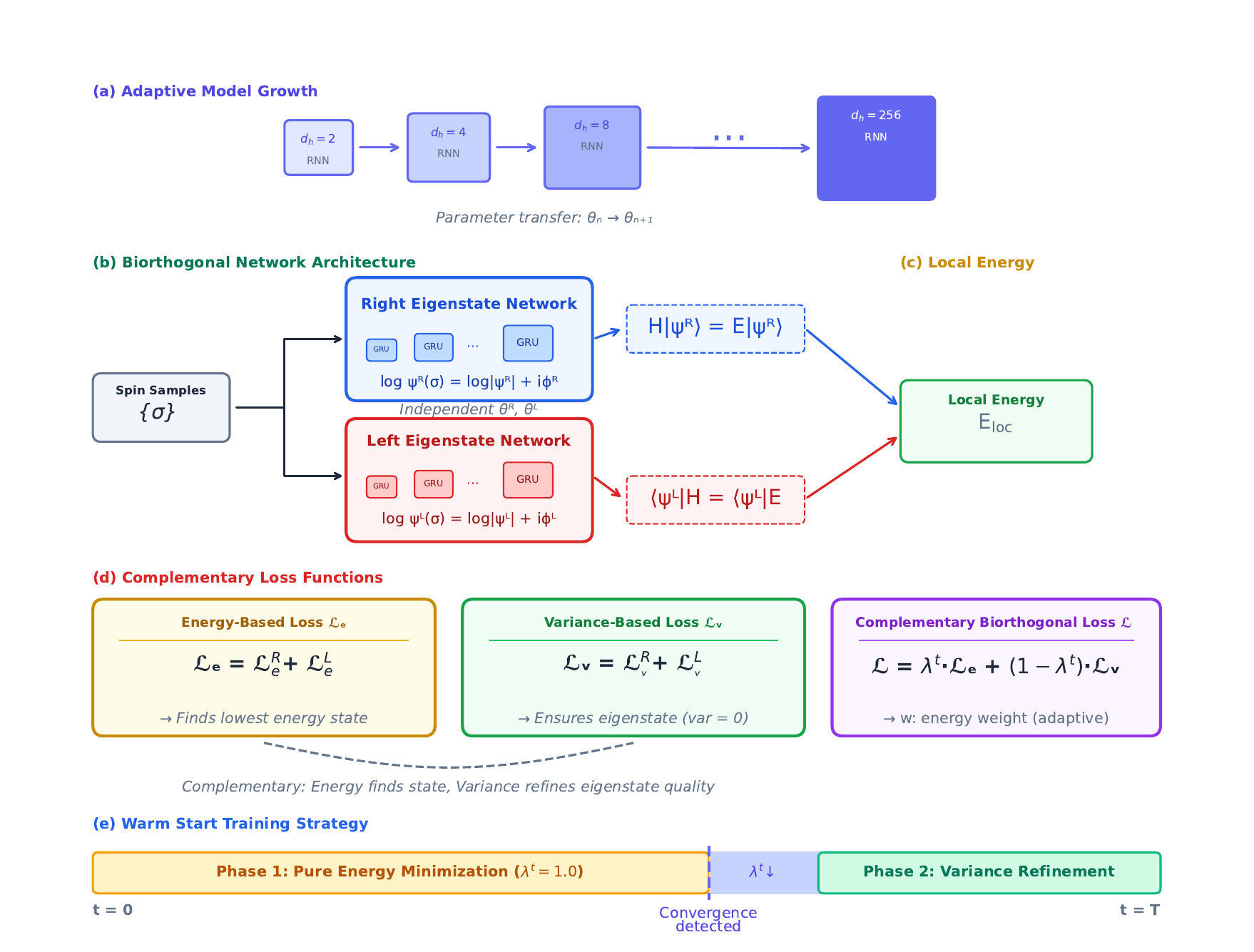}
    \caption{\textbf{Complementary optimization method for progressive adaptive state search}. We show a schematic presentation of our method. (a) We start by building an adaptive RNN, which progressively increases the network capacity during training following an adaptive scheme; then, (b) we initialize two independent RNNs with gated recurrent unit (GRU) cells to represent the right and left eigenstates of the model shown in blue and red, respectively. (c) We use the outputs from these RNNs to determine the local energy. As a next step, (d) we perform a complementary optimization in two phases: first we do an energy-based optimization to target the ground-state energy (shown in yellow), and second, we carry out a variance-based optimization to target the correct pair of biorthogonal eigenstates (in green). These two optimizations schemes are performed back and forth with an automatic adiabatic control of the energy weight parameter $\lambda^t$ (in purple). One cycle of this protocol is summarized in (e). In this figure $\{\sigma\}$ represents the short-hand notation for the spin set $|\bm{x}\rangle$ used in the main text.}
    \label{F1}
\end{figure*}

We present a complementary optimization method for the progressive adaptive state search for the variational and adiabatic determination of the ground-state properties of NH quantum systems, cf. Fig.~\ref{F1}. COMPASS employs two independent NQS autoregressive neural networks that represent the right and left eigenstates, which are trained using a combination of energy minimization, via a biorthogonal variational Monte Carlo (bVMC), and variance reduction. We propose a warm-start protocol that enables stable convergence, whereas the adaptive architecture automatically determines the required model's complexity based on the predefined schedule. Here, we define the ground state as the state with the smallest real part of the energy~\cite{wah2025}.

\textbf{\textit{1. Biorthogonal quantum mechanics.}} For NH Hamiltonians, the standard variational principle must be extended to a biorthogonal formalism \cite{RevModPhys.93.015005, Brody_2014}. Unlike Hermitian systems, where the left and right eigenstates are related by Hermitian conjugation, non-Hermitian systems require independent left and right eigenstates satisfying the biorthonormal condition
\begin{equation}\label{Eq3}
    \langle \Psi_{i}^L| \Psi_{j}^R\rangle =\delta_{i,j},
\end{equation}
where $i$ and $j$ are state indices. The energy expectation value takes the form
\begin{equation}\label{Eq4}
    \langle H\rangle = \frac{\langle \Psi_{i}^L|H|\Psi_{i}^R\rangle}{\langle \Psi_{i}^L| \Psi_{i}^R\rangle}.
\end{equation}

\textbf{\textit{2. Neural network ansatz.}} We parameterize both the right and left eigenstates using independent NQSs, cf. Fig.~\ref{F1}(b). Each NQS represents a complex-valued wavefunction in the computational basis $\{|\bm{x}\rangle\}$, with the spin configuration $|\bm{x}\rangle=\{x_1, x_2,...,x_N\}$ for a system of $N$ spins. The wavefunction amplitude $\Psi(\bm{x};\theta)$ for a spin configuration $|\bm{x}\rangle$ is factorized using the chain rule of probability (also known as conditional probability)
\begin{align}\label{Eq5}
\Psi(\bm{x};\theta)=\prod_{i=1}^N\Psi(x_i|x_1,x_2,...,x_{i-1};\theta),
\end{align}
where $\theta$ denotes the network parameters. Each conditional factor in Eq.~\eqref{Eq5} is parametrized as a complex number with separate amplitude and phase components
\begin{align}\label{Eq6}\ln{\Psi(\bm{s}_i|\bm{s}_{<i};\theta)}=\frac{1}{2}\ln{p(\bm{s}_i|\bm{s}_{<i})} +i\phi(\bm{s}_i|\bm{s}_{<i}),
\end{align}
where $\bm{s}_{i}$ is the one-hot encoding of ${x}_{i}$ (with $\bm{s}_0=0$),$\bm{s}_{1},\bm{s}_{2},...,\bm{s}_{i-1}$,  $p(s_i|s_{<i})$ is a normalized probability distribution, and $\phi(s_i|s_{<i})$ is the phase contribution. The log-amplitude is computed via softmax and the phase is bounded using softsign
\begin{align}\label{Eq7}
    \ln{p(\bm{s}_i=k|\bm{s}_{<i})} &=\ln\, \{\text{softmax}(\bm{o}_i^\textrm{Re})_k\}, \notag \\
    \phi(\bm{s}_i=k|\bm{s}_{<i})&=\pi \,\text{softsign}{(\bm{o}_{i,k}^\textrm{Im})},
\end{align}
where $\bm{o}_i^\textrm{Re}$ and $\bm{o}_i^\textrm{Im}$ are the real and imaginary part of outputs of the dense layers applied to the RNN hidden state at position $i$, respectively. This construction ensures that the conditional phases lie in $[-\pi,\pi]$, and the probabilities are properly normalized. The hidden state $H_i$ is updated recurrently
\begin{align}\label{Eq8}
    H_i= f(H_{i-1}, \bm{s}_{i-1};\theta),
\end{align}
where $f$ is the RNN cell (gated recurrent unit (GRU)~\cite{cho2014learning}, long short-term memory (LSTM)~\cite{6795963}, or vanilla RNN~\cite{medsker1999recurrent}). The outputs are computed as $\bm{o}_i^\textrm{Re} =\bm{W}^\textrm{Re}H_i+\bm{b}^\textrm{Re}$, and $\bm{o}_i^\textrm{Im} =\bm{W}^\textrm{Im}H_i+\bm{b}^\textrm{Im}$, where $\bm{W}^{n}$ and $\bm{b}^{n}$ are, respectively, weight and biases. The biorthogonal ansatz consists of two independent neural networks
\begin{align}\label{Eq9}
|\Psi^R(\theta^R)\rangle=\sum_{\bm{x}}\Psi^R(\bm{x};\theta^R)|\bm{x}\rangle, \notag \\
\langle\Psi^L(\theta^L)|=\sum_{\bm{x}}\langle\bm{x}|\Psi^L(\bm{x};\theta^L),
\end{align}
where each network has its own set of variational parameters $\theta^R$ and $\theta^L$ that are optimized independently.

\textbf{\textit{3. Biorthogonal VMC.}} In a NH setting, the traditional VMC, which is based on a pure estimator, does not hold anymore, and one must redefine what VMC means in this context. We employ the so-called bVMC, which unlike the traditional VMC, is defined as a mixed estimator [see Appendix~\ref{A1}] using both the left and right eigenstates.

\textbf{\textit{4. Complementary optimization.}} The model is optimized using two complementary loss functions, cf. Fig.~\ref{F1}(d). First, we construct a so-called biorthogonal ``energy-based'' loss function $\mathcal{L_e}$, whose aim is to solely find the lowest energy state (the ground state) based on the biorthogonal VMC such that $\mathcal{L_e}=\mathcal{L}^R_e+\mathcal{L}^L_e$, where $\mathcal{L}^R_e$ ($\mathcal{L}^L_e$) is the right (left) loss associated with each NQS. More details on their construction are provided in Appendix~\ref{A2}. Furthermore, we construct a ``variance-based'' loss function $\mathcal{L}_v$ by minimizing the energy variance. The aim of this loss function is to ensure that the state found is an eigenstate. This can be verified by knowing that each eigenstate of the Hamiltonian satisfies the zero-variance condition~\cite{solinas2025biorthogonal, zhang2025observation}. The resulting biorthogonal loss function is similarly defined as $\mathcal{L}_v=\mathcal{L}^R_v+\mathcal{L}^L_v$, where $\mathcal{L}^R_v$ $(\mathcal{L}^L_v)$ is the right (left) loss function computed from the right, and left energy variance $\sigma^2_R(\varepsilon)=(H^{\dagger}-\varepsilon^*)(H-\varepsilon),\, \sigma^2_L(\varepsilon)=(H-\varepsilon)(H^{\dagger}-\varepsilon^*)$, where $\varepsilon = \langle E_{\text{loc}}\rangle$ is an energy estimate that reduces the energy variance $\sigma^2=\mathbb{E}[|E_{\text{loc}}-\langle E_{\text{loc}}\rangle|^2]$. Therefore, for an eigenstate with eigenvalue $\varepsilon=E$, the variance is $\sigma^2=0$.

The complementary loss function used for optimization is then composed of the ``energy-based" loss function and the ``variance-based" loss function in the following way
\begin{align}\label{Eq11}
    \mathcal{L}=\lambda^t \mathcal{L}_e + (1-\lambda^t)\mathcal{L}_v,
\end{align}
where $\lambda^t \in [0,1]$.
Subsequently, in phase 1, the network is trained with pure energy loss (with energy weight $\lambda^t=1$) to rapidly approach the ground-state manifold. Training proceeds until the energy converges, which is detected by monitoring the change in energy over a sliding window $\Delta E =|E_t -E_{t-w}|< \epsilon_c$, where $w$ is the window size, $t$ is the current training steps, and $\epsilon_c$ is the convergence threshold. Upon convergence, the loss transitions automatically to variance minimization in phase 2, and the energy weight is adiabatically reduced using $\lambda^{t'} = \text{max} \big(\lambda_{\text{min}},\lambda^{t}(1-\gamma)\big)$, where $\gamma$ is the decay rate (see Fig.~\ref{F1}(e)), and $\lambda_{\text{min}}$ the minimum allowed value of the energy weight . This gradual transition prevents destabilization, while refining the eigenstate approximation. We refer to this procedure as the ``warm-start'' training strategy. This procedure is repeated a few times, while adiabatically adapting $\lambda^t$, so that one can smoothly switch back and forth with either loss functions. Thus, one ensures that the algorithm converges to the correct pair of biorthogonal states and that those states are the ones with the smallest energy (ground state energy).

An ablation study on the accuracy of the complementary approach compared to pure energy and pure variance minimization is presented in Appendix~\ref{A5}.

\textbf{\textit{5. Adaptive neural network.}} We used the adaptive framework proposed in Ref.~\cite{mcnaughton2025adaptive} with small modifications to speed up the entire process, cf. Fig.~\ref{F1}(a). Here, the network capacity progressively increases during training following an adaptive scheme. Similar to the transfer learning approach~\cite{zen2020transfer,wah2025}, we started with an RNN with a small hidden dimension $d_h=2$. The network is trained until convergence, and the parameters are transferred to a larger network with a hidden dimension $d_h \to 2d_h$ with no reset of $\lambda^t$. The transfer preserves learned weights, while initializing new parameters with small random values $\eta$~\cite{mcnaughton2025adaptive}
\begin{align}\label{Eq12}
\bm{W}_{\text{new}}=
 \begin{bmatrix}
\bm{W}_{\text{old}} & \eta \\
\eta & \eta
\end{bmatrix},
\end{align}
where $\eta \approx \mathcal{U}(-10^{-n}, 10^n)$ with $n$ being the growth stage index. This approach allows for the efficient exploration of the model capacity needed for accurate representation, while avoiding overfitting in the early training stages. We will later demonstrate that this adaptive framework is faster and more accurate than the static framework.

\textbf{\textit{6. Final energy estimation.}} After all of the above prescriptions are performed, one computes the final ground-state energy $E_0 =\sum_{\bm{x}}\bm{w}(\bm{x}) E_{\text{loc}}(\bm{x})$ [see Appendix~\ref{A6} for more details], where $\bm{w}(\bm{x})$ is the importance weight computed from the ratio $\Psi^L(\bm{x})/\Psi^R(\bm{x})$ (mismatch between the left and right eigenstates for a given spin configuration). Finally, we estimate the biorthogonal overlap fidelity as $|\langle\Psi^L|\Psi^R\rangle|^2 \approx \mathbb{E}_{\bm{x}\sim \rho(\bm{x})}\bm{w}(\bm{x})$ [see Appendix~\ref{A7} for more details], where the probability distribution $\rho(\bm{x})$ is used for importance sampling. This sampling is either performed using both ans\"atze or one of them [see Appendix~\ref{A1}]. In contrast to the work in Ref.~\cite{solinas2025biorthogonal}, which reported a sampling problem, we achieve a significantly low energy variance, suggesting that our approach captures the biorthogonal pair more accurately [see Appendix~\ref{A3}]. This improvement is due to the autoregressive nature of the NQS, which allows exact sampling. In addition, we achieve high (close to $1$) fidelity of the obtained eigenpair even with increasing system size [see Appendix~\ref{A7}], suggesting that for the given parameter space, our algorithm recovers the correct eigenpair. However, close to or at an exceptional point, this fidelity is expected to decrease due to the non-orthogonality of the eigenpair. Interestingly, our method enables the simulation of 1D NH systems up to $N=200$ and 2D NH systems up to $N=100$ spins, which at the best of our knowledge, is the largest system size reported to date in both dimensions.

\subsection{Comparison to other methods}

While a variance minimization approach for NH systems was recently used in Ref.~\cite{solinas2025biorthogonal} to find the ground state, our COMPASS approach fundamentally differs from it in several key aspects, offering significant computational and methodological advantages.

First, while Solinas et al.~\cite{solinas2025biorthogonal} rely on Markov Chain Monte Carlo (MCMC) \cite{geyer1992practical, gilks1995markov, melko2019restricted} sampling to generate configurations from the probability distribution $|\psi_R(\boldsymbol{\sigma})|^2$, our autoregressive RNN architecture enables exact, direct sampling, allowing configurations to be drawn sequentially without rejection \cite{medsker2001recurrent,yu2019review}. This eliminates several well-known limitations of MCMC methods: there is no burn-in period required, the samples are statistically independent, and autocorrelation effects are entirely absent \cite{geyer1992practical, brooks1998markov,carlo2004markov,gilks1995markov, melko2019restricted}. Consequently, the effective sample size equals the actual number of samples drawn, yielding gradient estimates with substantially lower variance \cite{melko2019restricted, geyer1992practical}. Furthermore, MCMC sampling can suffer from ergodicity issues near phase transitions or EPs, where the sampler may become trapped in metastable configurations~\cite{melko2019restricted, geyer1992practical, brooks1998markov,carlo2004markov}. Our direct-sampling approach circumvents this problem \cite{medsker2001recurrent, PhysRevResearch.2.023358}.

Second, while Wah et al.~\cite{wah2025} showed an alternative to bypass biorthogonality and reduce computational cost, we use an adaptive curriculum learning strategy that offers multiple advantages over their fixed-architecture approach. Small networks converge rapidly to capture the coarse features of the wavefunction, providing an effective form of regularization during early training \cite{mcnaughton2025adaptive}. The progressive introduction of complexity smooths the optimization landscape, and parameter transfer ensures that knowledge accumulated at smaller capacities initializes larger networks in favorable regions of the parameter space. This allows the simulation of larger system size within less computational time than previously reported in Refs.~\cite{wah2025,solinas2025biorthogonal}. Also, Ref.~\cite{solinas2025biorthogonal} employs pure variance minimization, which targets eigenstates without a bias toward any particular energy. While mathematically elegant, this approach can converge to excited states rather than the ground state, necessitating careful initialization strategies, such as adiabatic parameter sweeps from Hermitian limits as discussed in the paper. Our warm-start protocol first minimizes the energy to target the ground state, and then adiabatically introduces variance refinement, combining ground-state selectivity with eigenstate precision.

Finally, from a computational standpoint, our approach offers favorable scaling. Generating a single sample requires $\mathcal{O}(N)$ operations through the sequential autoregressive pass, whereas MCMC methods require $\mathcal{O}(N \times \tau_{\mathrm{burn}})$ operations, where $\tau_{\mathrm{burn}}$ is the burn-in time, which itself can scale unfavorably near critical points. Additionally, the independence of the samples enables straightforward parallelization across batch dimensions without concerns regarding chain mixing.

\begin{figure}
    \centering
    \includegraphics[width=0.49\textwidth]{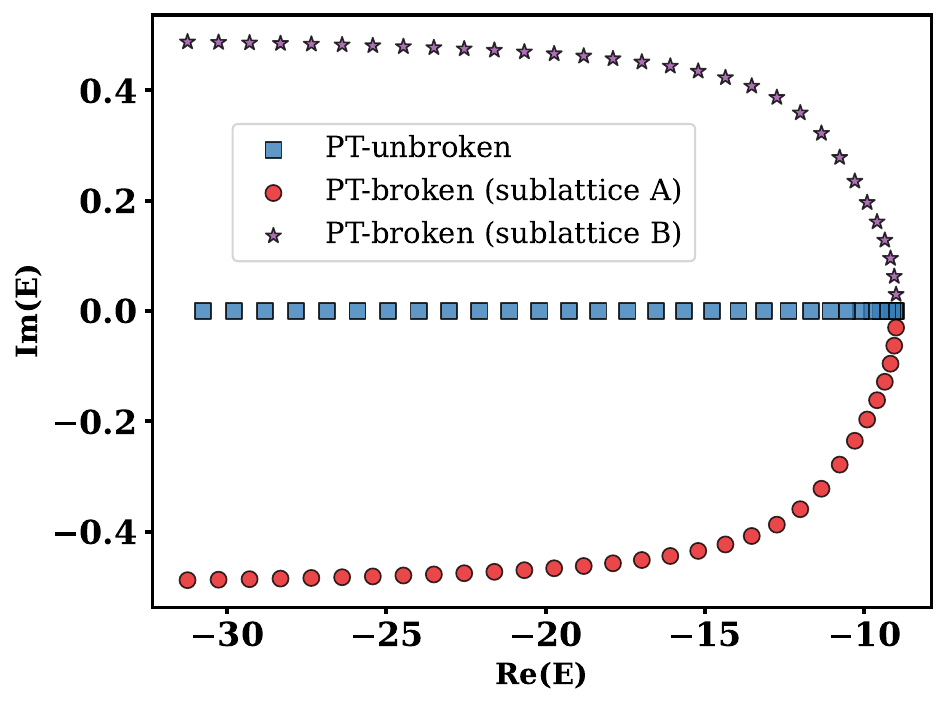}
    \caption{\textbf{Broken vs unbroken $PT$ symmetry}. We plot the imaginary part of the ground-state energy of the 1D Ising chain against its real part for different values of $\eta \in[0,3]$, and $\xi=\eta/10$. We consider $\delta \lambda =0.1+i0.1$ when we perturb sublattice $A$, and $\delta \lambda =0.1-i0.1$ when we perturb sublattice $B$. Open boundries conditions (OBCs) are considered, and $N=10$.}
    \label{F2}
\end{figure}

\section{Architectural Bias in Biorthogonal Adaptive Neural Quantum States for 1D models}\label{S2}

In this section, Using COMPASS, we show that ansatz choice for non-Hermitian systems is physically decisive rather than a matter of convenience. We also motivate our choice of an adaptive NQS.

\subsection{1D Models} \label{sec:meth_1D_ex}

To apply the above method, we study the ground-state properties of a 1D quantum transverse-field Ising chain (TFIM) of $N$ spin-$1/2$ particles on a lattice subject to a complex staggered magnetic field. The NH transverse field is chosen such that the system is parity-time (PT) symmetric. The Hamiltonian describing such a system is given by
\begin{align}\label{Eq1}
    H = -J \sum_{j=1}^N \sigma_j^z \sigma_{j+1}^z -g \sum_{j \in A}\sigma_j^x-g^* \sum_{j \in B}\sigma_j^x,
\end{align}
where $g= \eta + i \xi$ is applied to the spins on sublattice $A$, and $g^*= \eta - i \xi$ to the spins on sublattice $B$ with $\eta, \xi \in \mathbb{R}$~\cite{wah2025}. $\sigma_j^z$ and $\sigma_j^x$ are Pauli matrices. Spins are coupled via a spin coupling $J$, and only nearest-neighbor coupling is considered. Unless stated otherwise, we will work in the regime where $J=1$. 

One can break the PT symmetry of this Hamiltonian by adding a small perturbation $\delta \lambda = \delta \alpha +i \delta \beta$ (see Fig.~\ref{F2}) on either the $A$ or $B$ sublattice such that $g \rightarrow g + \delta \lambda$ or $g^* \rightarrow g^* + \delta \lambda^*$, respectively, in the Hamiltonian in Eq.~\eqref{Eq1}.
Depending on whether the perturbation is applied to sublattices $A$ or $B$, the system can either undergo a loss process (red circles) or a gain process (pink stars), respectively. In the absence of this perturbation, the gain and loss is balanced, and the system is $PT$-symmetric (blue square). All the parameters required to simulate this system are listed in Appendix~\ref{A4}.

\begin{table*}[!ht]
\centering
\begin{tabular}{|c|c|c|c|c|c|}
\hline
Size & Method/Ansatz & $\text{Re}(E_0)/N$ & $\text{Im}(E_0)/N$ & $PT$-symmetric & Time (hh:mm:ss) \\ \hline
 & cRNN (adaptive) & $-1.743438795651$ & $-4.126049 \times 10^{-6}$ & True & $00:01:04$ \\
& pRNN (adaptive) & $-1.743264567199$ & $\mathbf{2.45171 \times 10^{-7}}$ & True & $00:01:00$ \\
$N=10$ & ED & $-1.743387214507$ & $2.25796 \times 10^{-16}$ & True & $00:02:10$ \\
& cRNN (static) & $-1.743215992579$ & $-6.490997 \times 10^{-6}$ & True & $00:02:00$ \\
& pRNN (static) & $-1.743696909127$ & $\mathbf{1.94177 \times 10^{-7}}$ & True & $00:03:13$ \\
\hline
& cRNN (adaptive) & $-1.733616932162$ & $9.71826168 \times 10^{-4}$ & False & $01:08:00$ \\
& pRNN (adaptive) & $-1.742009430969$ & $\mathbf{2.300877 \times 10^{-6}}$ & True & $01:00:00$ \\
$N=100$ & SE & $-1.743750374535$ & $2.00011 \times 10^{-16}$ & True & $-$ \\
& cRNN (static) & $-1.734498442123$ & $6.71078351 \times 10^{-4}$ & False & $03:40:12$ \\
& pRNN (static) & $-1.745085781971$ & $\mathbf{2.299659 \times 10^{-6}}$ & True & $03:18:00$ \\
\hline
\end{tabular}
\caption{\textbf{Adaptive vs static neural quantum state.} This table represents the ground-state energies per spin of the $PT$-symmetric TFIM with some symmetry detection layer added on the ansatz. We see that depending on which parameter regime we simulate the choice of the ansatz is crucial. We also show that the adaptive RNN is computationally more efficient than its static version. }
\label{tab1}
\end{table*}

\subsection{ Adaptive vs static neural quantum states}

In this subsection, we compare the performances of the adaptive and static RNN for the PT-symmetric model in Eq.~\eqref{Eq1}. For the static version, we consider an RNN (GRU) with a hidden dimension $\bm{d}_h=256$. 

We simulate the ground-state energy of the model for $N=10$ and $N=100$ spins, and record the time for all simulations in Tab.~\ref{tab1}. For both the static and adaptive RNN, we use a complex ansatz (cRNN) and a real (positive) ansatz (pRNN). We find that the adaptive RNN runs faster than its static counterpart for both system sizes, where the speed up becomes more significant for larger system sizes. Indeed, for $N=100$, we observe that the adaptive RNN runs at least three times faster than its static version. This suggests that the adaptive framework is less time consuming and computationally less expensive.

\subsection{Ansatz induced spontaneous symmetry breaking in non-Hermitian systems}

Having established that adaptive RNNs outperform their static counterparts, we will hereafter exclusively employ the adaptive framework. In Hermitian systems, any quantum state can in principle be represented using a complex-valued wavefunction. However, in the NH setting this generalization is more nuanced. While complex ans\"atze constitute the most general form of wavefunction representation, NH systems inherently involve complex-valued spectra and nontrivial symmetries, which introduce additional subtleties in the choice of ansatz used to represent Hamiltonian eigenstates. In this study, we argue that the ansatz in NH systems must be selected with greater care, as the use of a formally general ansatz does not guarantee an accurate approximation. In particular, we demonstrate that in certain cases, an inappropriate choice of ansatz can lead to spontaneous symmetry breaking \cite{PhysRevD.10.500, beekman2019introduction}, as summarized in Table~\ref{tab1}.

For the $PT$-symmetric model \eqref{Eq1}, we simulate the ground state using both  real and complex ans\"atze (see Table~\ref{tab1}). We augment the NQS with a symmetry-detection layer that checks whether the wavefunction satisfies $PT$ symmetry. Specifically, we verify whether $\Psi_R(\bm{x}) = \bigl( \Psi_L (\bm{x})\bigr)^{*} + \text{offset}$, with $\text{offset} = 10^{-10}$. Equivalently, one may check whether the phases (that is, the imaginary parts) are sufficiently small to be neglected.

\begin{figure*}[ht!]
    \centering
    \subfloat[\centering ]{{\includegraphics[width=0.49\textwidth]{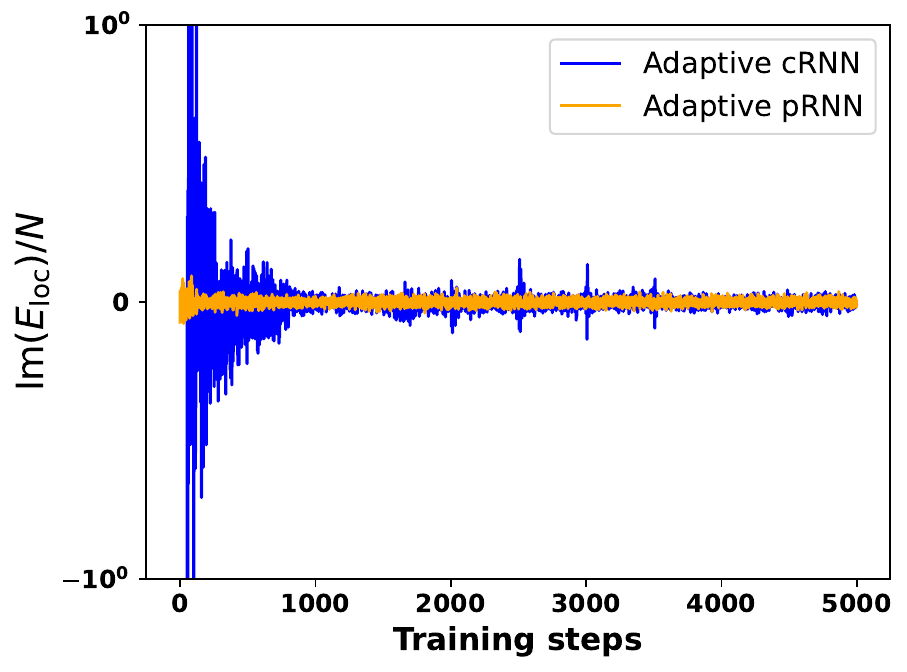} }}
    \subfloat[\centering ]{{\includegraphics[width=0.49\textwidth]{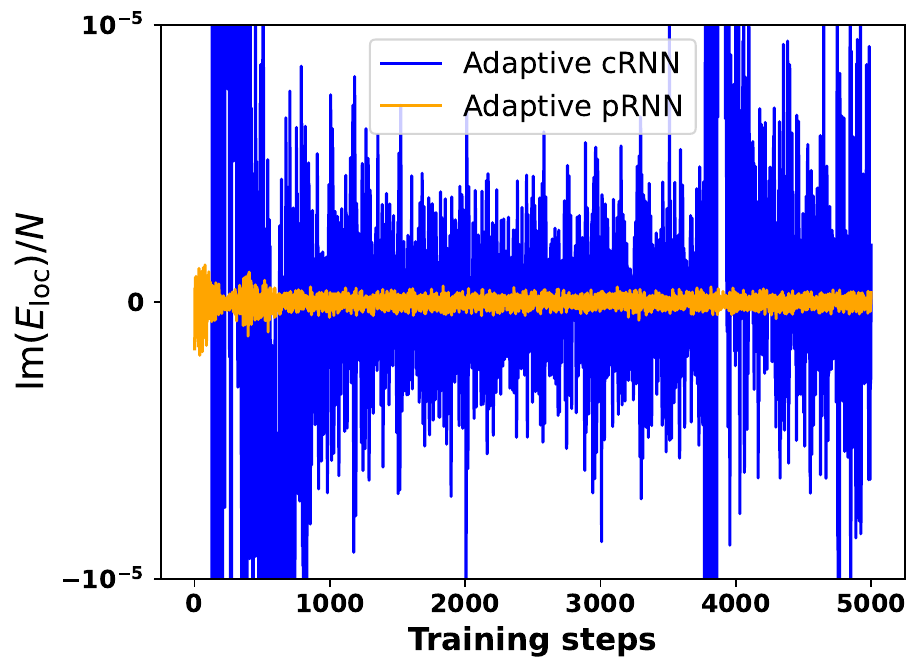} }}
    
    \caption{\textbf{Energy per spin for the 1D $PT$-symmetric Hamiltonian}. We plot the imaginary part of the ground-state energy per spin of the 1D $PT$-symmetric TFIM in Eq.~\eqref{Eq1}, for \textbf{(a)} $N=10$, and \textbf{(b)} $N=100$ spins using both the adaptive cRNN (in blue) and pRNN (orange). We consider $J=1$, $\eta=1.6$ and $\xi=0.16$.}
    \label{F3}
\end{figure*}

For $N=10$ spins, the NQS successfully finds the ground state of the model, and the symmetry-detection layer confirms that the wavefunction is $PT$-symmetric. However, in this regime, we observe that the pRNN yields an imaginary part of the energy, which is one order of magnitude smaller than that obtained with the cRNN (see Fig.~\ref{F3}(b)). For $N=100$, the ground state is still expected to be $PT$-symmetric, as shown by the series expansion methods~\cite{lenke2021high, wah2025}. Nevertheless, only the pRNN ansatz can approximate the ground-state energy with a sufficiently small imaginary part (see Fig.~\ref{F3}(b)), while the symmetry-detection layer indicates that the resulting ground-state wavefunction is no longer $PT$-symmetric for the cRNN (see Table~\ref{tab1}). Thus, although the unconstrained complex ansatz is formally the most general, it fails to represent the correct state of the $PT$-symmetric system. This naturally raises the question of what happens for a generic NH Hamiltonian.

\begin{figure*}[ht!]
    \centering
    \subfloat[\centering ]{{\includegraphics[width=0.48\textwidth]{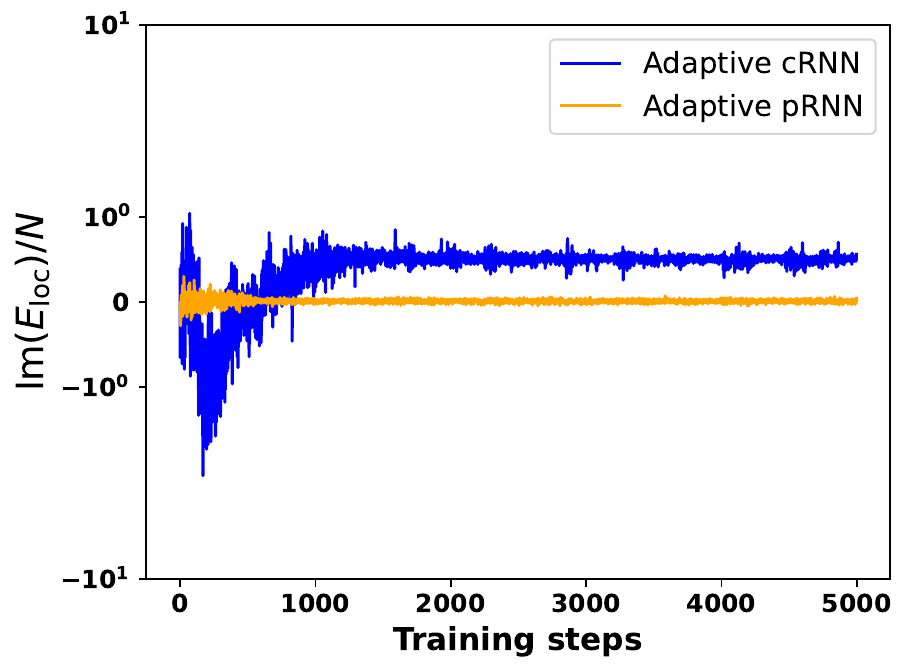} }}
    \subfloat[\centering ]{{\includegraphics[width=0.48\textwidth]{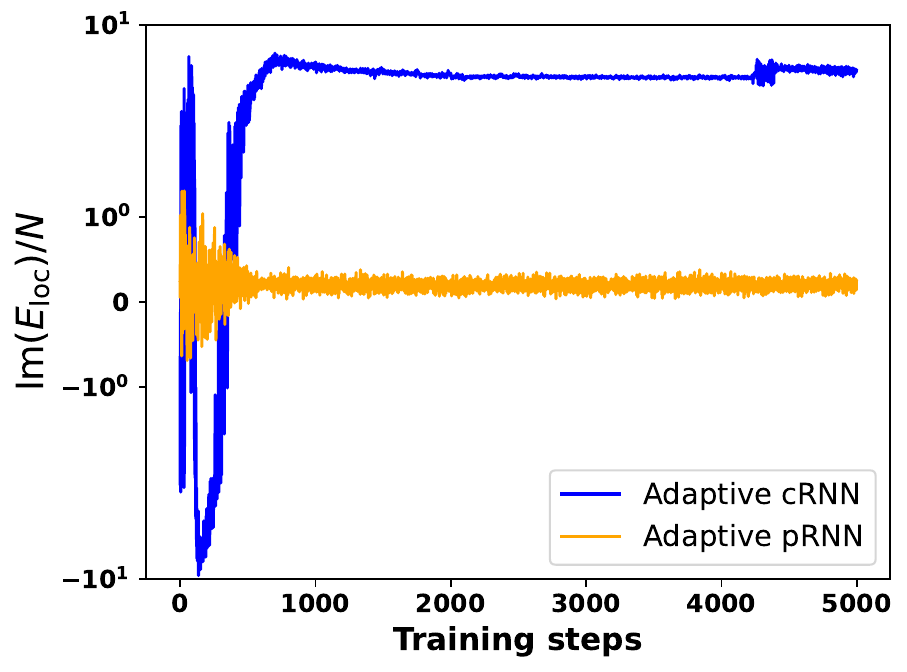} }}
    
    \caption{\textbf{Energy per spin for the generic 1D non-Hermitian Hamiltonian}. We plot the imaginary part of the ground-state energy per spin of the generic 1D TFIM in Eq.~\eqref{Eq2}, for  \textbf{(a)} $N=10$, and \textbf{(b)} $N=100$ spins using both the adaptive cRNN (in blue) and pRNN (orange). We consider $J=1$, $\eta=1.6$, $\xi=0.16$ and $\delta \lambda = ?$. }
    \label{F4}
\end{figure*}

To address this, we consider a generic NH Hamiltonian of the form discussed in Sect.~\ref{sec:meth_1D_ex} with a complex magnetic field on the $A$ sublattices, i.e.,
\begin{align}\label{Eq2}
    H = -J \sum_{j=1}^N \sigma_j^z \sigma_{j+1}^z -(g+\delta \lambda) \sum_{j \in A}\sigma_j^x-g^* \sum_{j \in B}\sigma_j^x,
\end{align}
and perform the same simulations. For $N=10$, the pRNN underestimates the imaginary part of the ground-state energy, as the NQS is biased toward a predominantly real representation and, therefore, tends to neglect the phase information. Consequently, the estimated ground-state energy becomes nearly real rather than complex (see Fig.~\ref{F4}(a)). In contrast, the cRNN recovers a more accurate complex ground-state energy. For $N=100$, the discrepancy between the imaginary parts of the ground-state energy estimated using a cRNN and a pRNN becomes even more pronounced (see Fig.~\ref{F4}(b)).

Overall, for NH systems, a more expressive ansatz is not always the optimal choice. In principle, a sufficiently expressive complex ansatz can represent any wavefunction, including those that are real or obey specific symmetry constraints. However, due to the highly nontrivial optimization landscape of NH systems, increased expressivity can in practice bias the optimization process. In particular, for $PT$-symmetric systems, the additional degrees of freedom present in a complex RNN introduce directions in the parameter space that do not improve $\mathrm{Re}(E)$, but allow $\mathrm{Im}(E)$ to grow without incurring any penalty within standard variational optimization schemes. We therefore argue that for $PT$-symmetric systems in the unbroken phase, a real ansatz formulated in an appropriate basis provides a more suitable trial wavefunction, as it constrains the optimization to the correct physical manifold.

In contrast, for generic NH Hamiltonians with generally complex spectra, a complex ansatz is essential to faithfully capture the system's properties and avoid  the systematic underestimation of complex observables. We note, however, that in certain cases a complex ansatz may be explicitly constrained to represent spontaneously unbroken $PT$-symmetric states. This can be achieved by imposing symmetry constraints on the ansatz prior to optimization, at the cost of an increased architectural complexity and computational overhead. Overall, when the eigenstate has a real eigenvalue or possesses a specific structural property, employing an ansatz that ``matches'' this structure is practically superior, even if a ``more general'' ansatz is, in principle, sufficiently expressive.

\begin{figure*}[ht!]
    \centering
    \subfloat[\centering ]{{\includegraphics[width=0.40\textwidth]{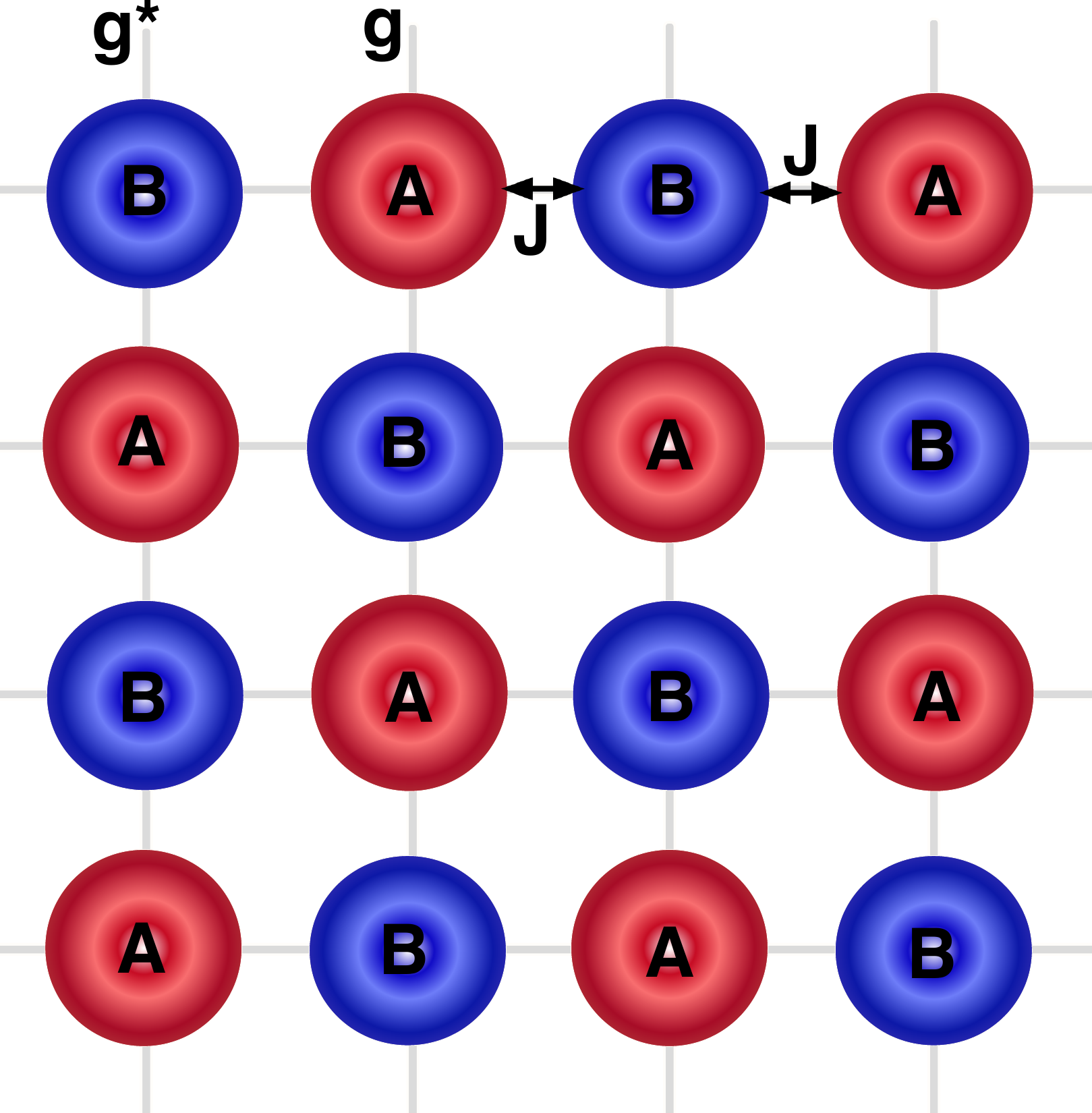} }}
    \subfloat[\centering ]{{\includegraphics[width=0.55\textwidth]{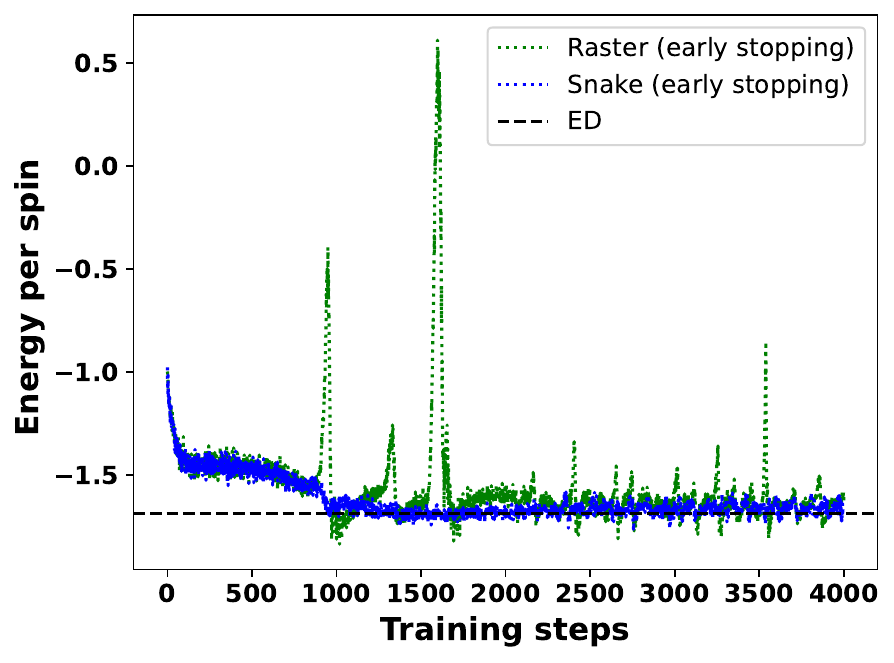} }}
    
    \caption{\textbf{The 2D $PT$-symmetric TFIM}. We present \textbf{(a)} a schematic representation of the 2D $PT$-symmetric TFIM \eqref{eq:2D_TFIM}, and \textbf{(b)}~plot the ground state energy per spin with early stopping with the snake scan (blue), and the raster scan (green). We compare the results to ED (black). We consider OBCs, $J=1$, $\eta=1.6$, $\xi=\eta/10$, $L_x \times L_y = 3 \times 3$, a patience step $N_{\text{pat}}=500$, and a convergence threshold of $10^{-5}$. The snake scanning is more stable and reliable compared to the raster scanning. ED could not simulate $L_x \times L_y = 4 \times 4$.}
    \label{F5}
\end{figure*}

\section{Extension to a Two-dimensional Model}\label{S3}

\subsection{Training}
We showed in the previous section that our method works for 1D models. Now, we extend our study to 2D models, and focus on a $PT$-symmetric TFIM [see Fig.~\ref{F5}(a)]. The model is a 2D extension of the Hamiltonian in Eq.~\eqref{Eq1} and can be written as
\begin{align}
    H = -J \sum_{\langle i,j \rangle}^N \sigma_i^z \sigma_{j}^z -g \sum_{i \in A}\sigma_i^x-g^* \sum_{i \in B}\sigma_i^x, \label{eq:2D_TFIM}
\end{align}
where $\langle.\rangle$ represents nearest-neighbor interactions along both directions.

Applying a 1D autoregressive RNN to 2D quantum lattice models requires serializing the 2D spin configuration into a 1D sequence, a step that has a nontrivial impact on the quality of the variational ansatz. Two natural site orderings exist for an $L_x\times L_y$ square lattice: the raster or row-major scan~\cite{raster1977179,PhysRevResearch.2.023358,van2016pixel} (where every row is traversed left-to-right giving flat index $k=\text{row}\times L_y +\text{column}$), and the snake or boustrophedon scan~\cite{snake2015scanning} (where rows are traversed in reverse directions) [see more details in Appendix.~\ref{A6}]. Both scanning techniques were discussed in the context of the RNN wavefunction for Hermitian systems in Ref.~\cite{PhysRevResearch.2.023358}, where raster scanning was used for 1D RNNs applied to Hermitian 2D lattices, and a zigzag path was introduced for the tensorized 2D RNN extension (which we leave for future work). As noted in that work and further discussed in Ref.~\cite{hibat2022supplementing}, when 1D RNNs are applied to 2D systems, the 2D structure of the lattice is not explicitly encoded, and some physically neighboring sites are inevitably separated in the network's sequential path. However, in the case of our model, we argue that the snake scan provides a systematic advantage over the raster that goes beyond this general consideration, specially at row boundaries where the sublattice symmetry of the model is preserved [see more details in Appendix.~\ref{A6}].

While our 1D implementation converges reliably within the allocated steps, cf. Figs.~\ref{F3} and \ref{F4}, the extension to 2D lattices introduces new computational challenges in the sense that the simulations are expected to take much longer than in the 1D case. This motivates the use of early stopping~\cite{PRECHELT1998761, prechelt2002early}, which is a tool used to prevent the overfitting of data. We monitor the real part of the energy and trigger early stopping if no improvement exceeding $10^{-5}$ is observed over $N_{\text{pat}}=500$ (patience) consecutive steps [see more details in Appendix.~\ref{A6}]. This threshold and the patience steps were chosen to balance the convergence quality with computational efficiency. Using this, we then find that our data converges within $4000$ training steps: We depict the ground-state energy per spin of the model with raster (in green) and snake (blue) scanning compared to ED for $L_x \times L_y = 3 \times 3$ in Fig.~\ref{F5}(b). A smooth convergence with snake scanning was observed, and we argue that this observation is first due to the symmetry of the model, which imposes a $A/B$ pattern that is broken by the raster configuration. In addition, the raster scanning introduces unphysical long-range interactions between two rows, which subsequently introduces some errors in the results since only nearest-neighbor interactions are included in the Hamiltonian [see more details in Appendix.~\ref{A6}]. We achieve relatively low variance with both scans, but the variance converges faster and smoothly with the snake scan as depicted in Appendix~\ref{A3}. We conclude that the COMPASS approach is generalizable to 2D systems, and we demonstrate its scalability by simulating $L_x \times L_y = 10 \times 10$ [see Fig.~\ref{F62}, and Appendices~\ref{A7}]. The parameters used to simulate this model, as well as the 1D models are listed in Appendix~\ref{A4}.

\subsection{Final energy}
\begin{figure}
    \centering
    \includegraphics[width=0.48\textwidth]{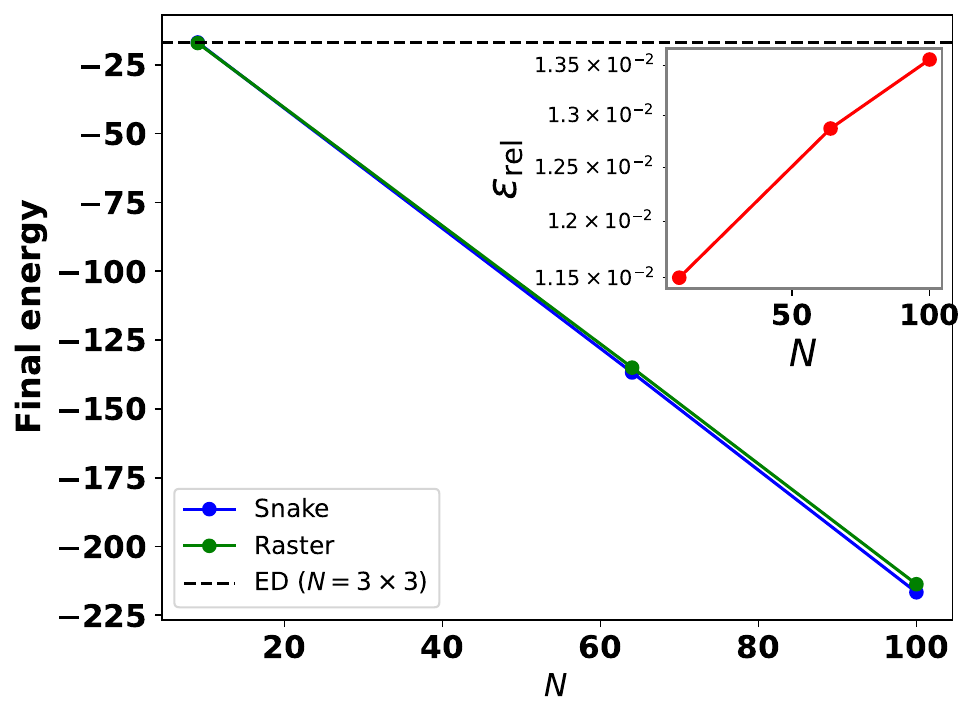}
    \caption{\textbf{Final energy for the 2D $PT$-symmetric TFIM model}. We plot the ground state energy of the 2D $PT$-symmetric TFIM \eqref{eq:2D_TFIM} computed after optimization for the snake scan (in blue) and raster scan (green) as a function of $N = L_x \times L_y$ for $L_x = L_y = 3, 8 , 10$, $J=1$, $\eta=1.6$, and $\xi=\eta/10$. In the inset, we show the relative error $\varepsilon_{\text{rel}}=|E_{\text{snake}}-E_{\text{raster}}|/|E_{\text{snake}}|$ and find that the snake scan yields an energy, which is two orders of magnitude more accurate than the energy obtained with the raster scan. OBCs are considered.}
    \label{F62}
\end{figure}

Here we plot the final energy for systems size up to $L_x \times L_y=10\times10$ after the training is completed. We observe that the energy obtained with the snake scan remains consistently lower than the one obtained with the raster scan (see Fig.~\ref{F62}). Note that this final energy is expected to be more accurate than the on-the-fly energy in Fig.~\ref{F5}, which is evaluated at intermediate parameter states during optimization, where the variational ansatz has not fully converged to the true ground state, and is therefore prone to noise. In contrast, the final energy is computed after the model has converged and settled into the minimum. 

In the inset of Fig.~\ref{F62}, we show the relative error between the final energies of both scanning methods. We find that $\varepsilon_{\text{rel}}$ is of order of $\approx 10^{-2}$ indicating that the snake scan is at least two orders of magnitude lower than the raster scan, which according to our previous argument suggest that snake scan is more accurate than the raster scan.

\section{Beyond stoquasticity: non-Hermitian control of frustration in the $J_1-J_2$ chain}\label{S4}

Let us now go beyond stoquastic models, and explicitly include frustration.
Here, we present study of a NH extension of the 1D spin-$1/2$ $J_1-J_2$ Heisenberg model, which, in the Hermitian limit and provided positive couplings, has a ground state endowed with a sign structure in the computational $z$ basis~\cite{PhysRevResearch.2.023358}. We reveal how non-Hermiticity reshapes the frustration in the chain, thereby providing the first study of a non-Hermitian model beyond stoquasticity.

\begin{figure*}[ht!]
    \centering
    \subfloat[\centering ]{{\includegraphics[width=0.33\textwidth]{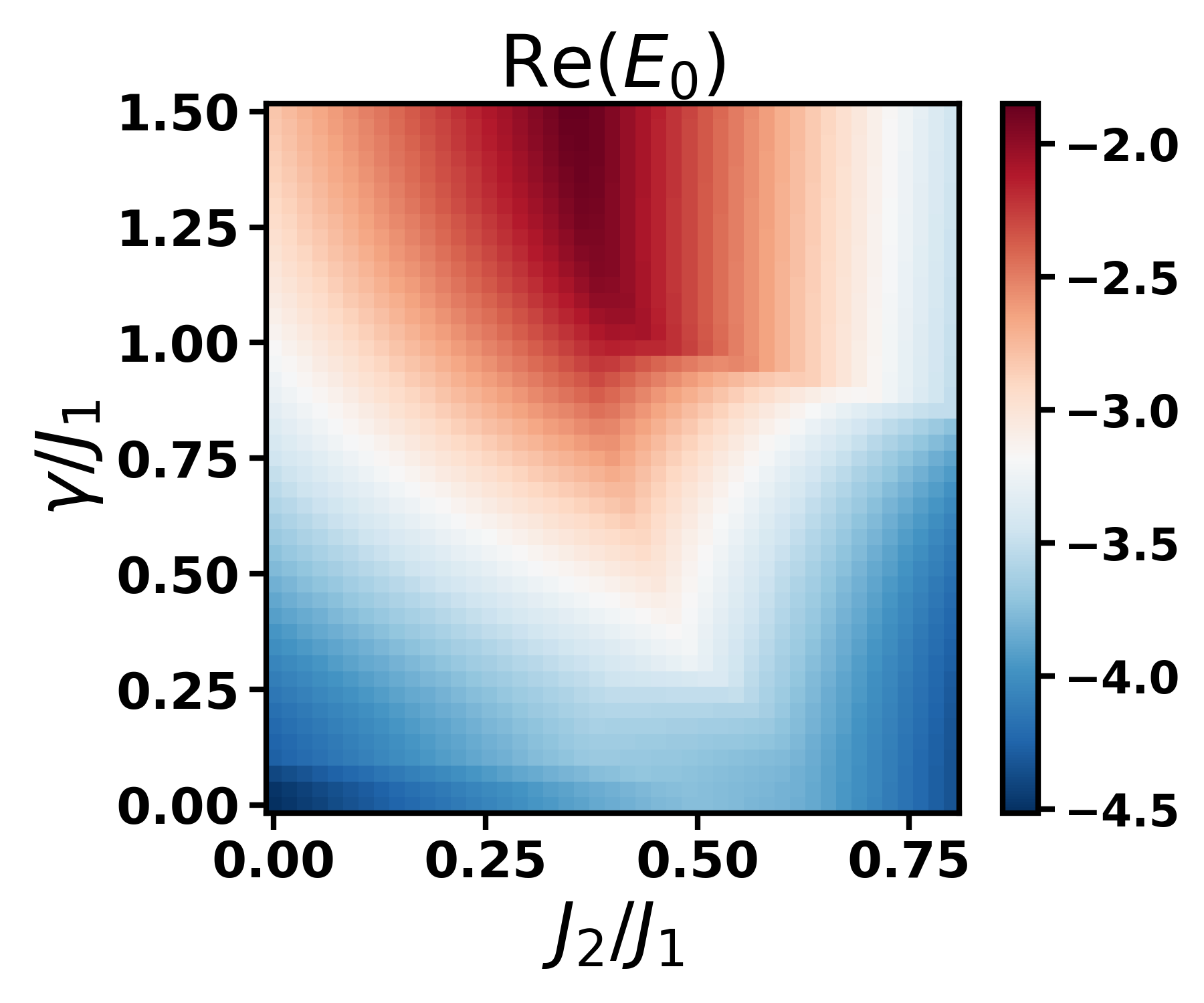} }}
    \subfloat[\centering ]{{\includegraphics[width=0.33\textwidth]{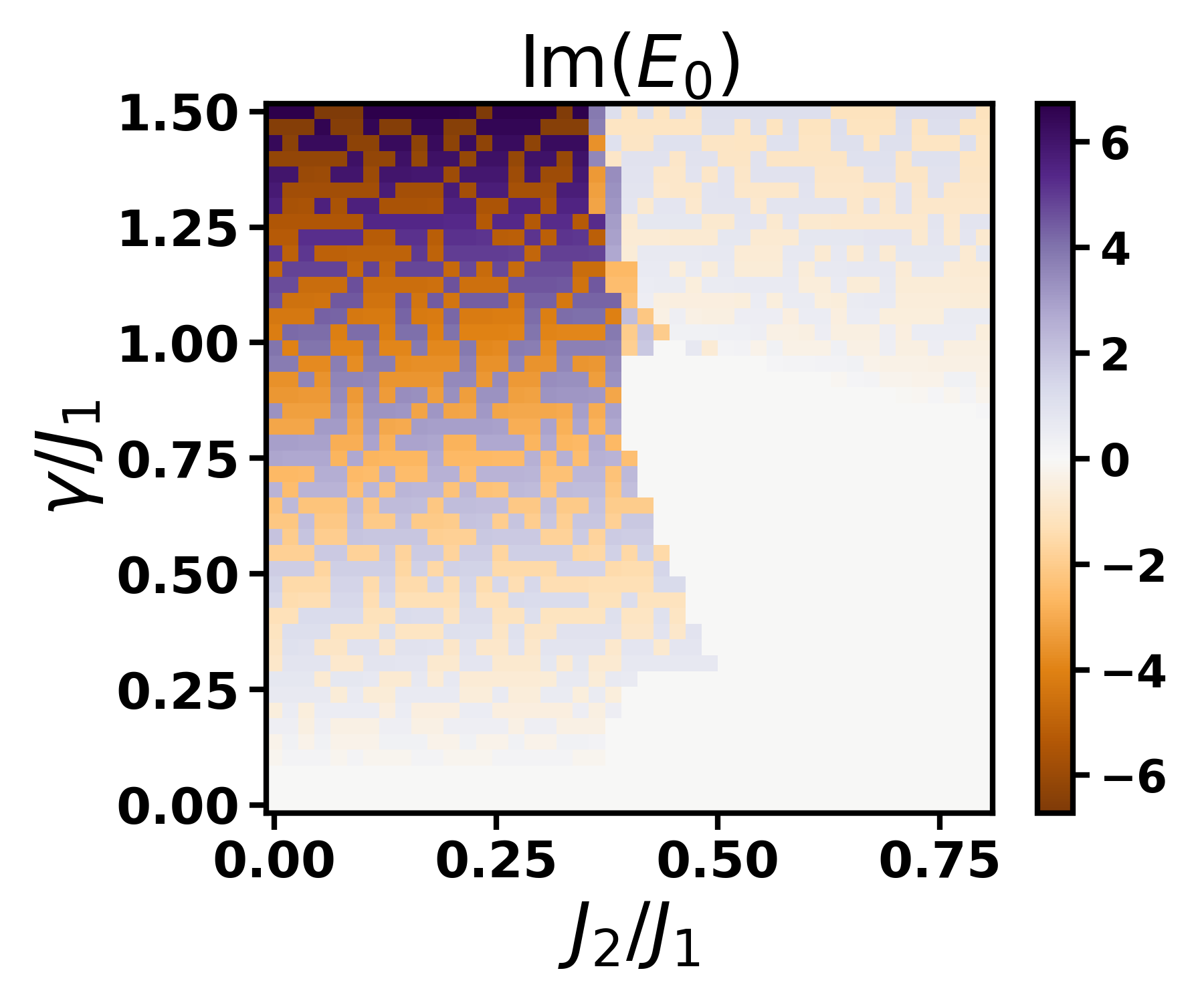} }}
    \subfloat[\centering ]{{\includegraphics[width=0.33\textwidth]{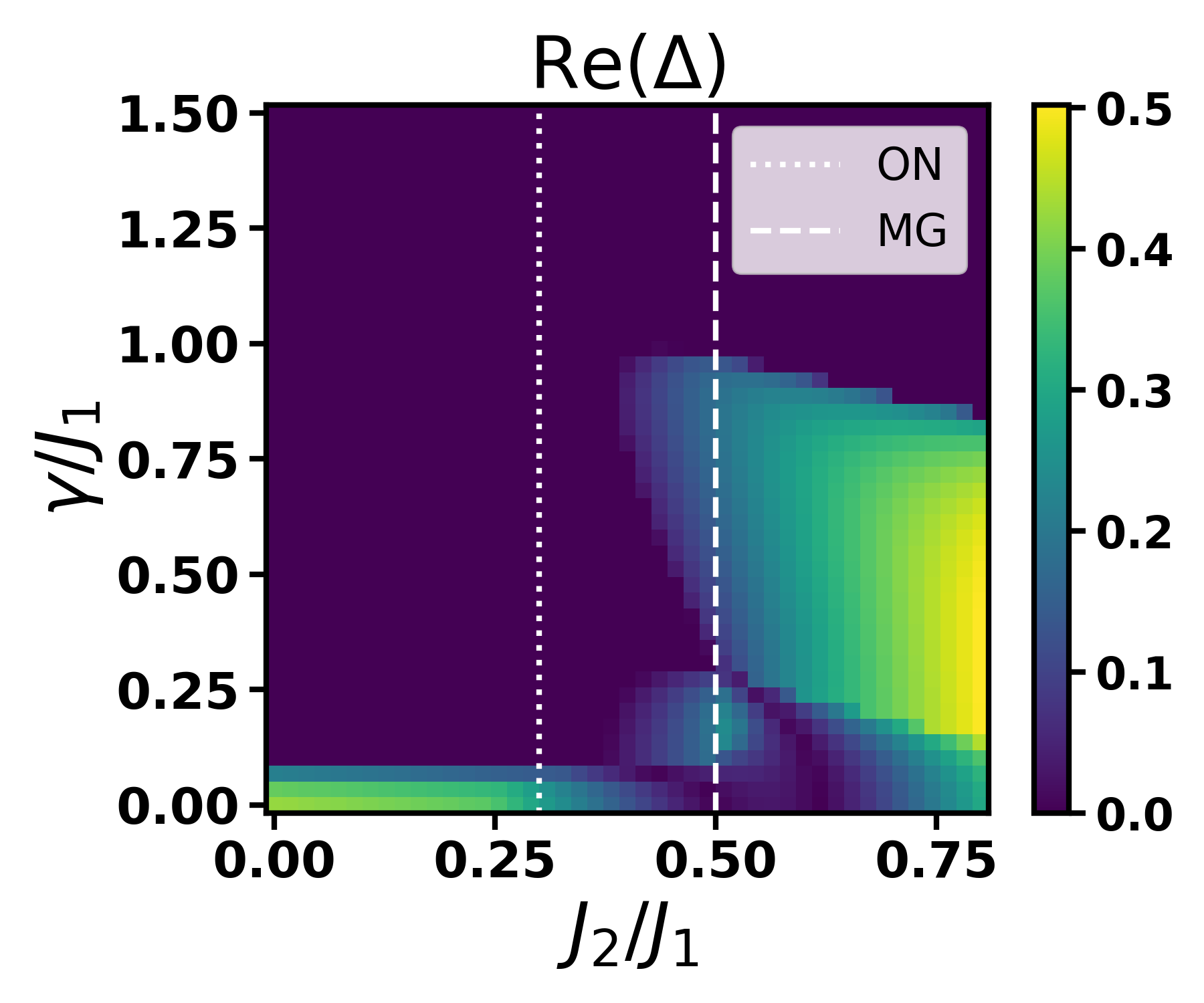} }}
    
    \caption{\textbf{Phase diagram of the NH $J_1-J_2$ with staggered magnetic field}. We show the phase diagram of the Hamiltonian in Eq.~\eqref{Eq13} using ED and presenting how the \textbf{(a)} real part, and the \textbf{(b)} imaginary part of the ground-state energy $E_0$, as well as the \textbf{(c)} real part of the energy gap $\Delta=E_1-E_0$ changes with non-Hermiticity $\gamma/J_1$, and the frustration parameter $J_2/J_1$. $E_1$ is the first excited state energy obtained from block-diagonalization. The dotted line (ON) is the Okamoto-Nomura point representing here the transition from the gaped to the gapless phase (since $E_0$ and $E_1$ are computed within the same sector $S^z=0$), and the dashed line (MG) is the Majumdar-Ghosh point where the ground state is exactly twofold degenerate. We considered $N=10$, $J_1=1$ and PBCs. We observe a gap protection of the $PT$-symmetry, and the lift of the MG degeneracy. }
    \label{F6}
\end{figure*}

\subsection{Models}

We study two NH Hamiltonians on a 1D chain of $N$ spin-$1/2$ degrees of freedom with PBCs.

\textbf{\textit{Model 1: Staggered imaginary field $J_1-J_2$}}. The first model we consider is given by the following
\begin{align}\label{Eq13}
        H_1 = J_1  \sum_{\left<i,j\right>}  S_i \cdot S_j
      + J_2 \sum_{\left<\left<i,j\right>\right>} S_i \cdot S_j
      + i\gamma  \sum_j (-1)^j  S_j^z,
\end{align}
where the first two terms constitute the standard Hermitian $J_1-J_2$ Heisenberg exchange over nearest-neighbor (NN, $\langle . \rangle$) and the next-nearest-neighbor (NNN, $\langle \langle . \rangle \rangle$) pairs. The third term is a staggered imaginary longitudinal field with strength $\gamma$, which is chosen such that Eq.~\eqref{Eq13} is $PT$-symmetric for even $N$. We also set $\gamma, J_1, J_2 \in \mathbb{R}$, and $S_i=(S_i^x,S_i^y,S_j^z)$ is a vector of quantum spin operator representing the angular momentum operators.

This Hamiltonian arises from the quantum trajectory (quantum jump) approach to open quantum systems. Consider a Lindblad master equation with staggered loss and gain operator operators such that, the spin-down decay $L_k=\sqrt{\gamma_k} \sigma_k^-$ acts on even site of the sublattice and the spin-up decay $L_k=\sqrt{\gamma_k} \sigma_k^+$ acts on odd sites. The effective NH Hamiltonian governing the evolution conditioned on zero quantum jumps contains the term $-\frac{i\gamma}{2}\sum_{k\,\text{even}}\frac{1+\sigma_k^z}{2} 
- \frac{i\gamma}{2}\sum_{k\,\text{odd}}\frac{1-\sigma_k^z}{2}$, where we used 
$L_k^\dagger L_k = \sigma_k^+\sigma_k^- = \frac{1+\sigma_k^z}{2}$ on even sites 
and $L_k^\dagger L_k = \sigma_k^-\sigma_k^+ = \frac{1-\sigma_k^z}{2}$ on odd sites. 
Dropping the uniform imaginary constant $-\frac{i\gamma}{4}\sum_k \mathbf{1}$, 
which carries no physical consequence, this reduces to the staggered imaginary 
onsite field $-\frac{i\gamma}{4}\sum_k (-1)^k \sigma_k^z = -\frac{i\gamma}{2}\sum_k (-1)^k S_k^z$. This construction is well defined and experimentally accessible, where the protocol will be to prepare the system, post-select on no decay events, and the governing dynamics will be governed by a NH Hamiltonian \cite{tpfcn3bq, PhysRevX.4.041001}.

This model is also connected to the Yang-Lee edge singularity through the Ising model with an imaginary field~\cite{PhysRevLett.40.1610, GvonGehlen_1991}, and the construction can in principle find applications in trapped ions, cavity QED, and atoms in optical lattices \cite{PhysRevX.4.041001}.  Crucially, $H_1$ conserves the total $S_{\text{tot}}^z =\sum_i S_i^z$, since both the Heisenberg exchange and the diagonal field commute with $S_{\text{tot}}^z$. This allows block-diagonalization [See Appendix~\ref{A8} for more details] in the $S^z=0$ sector of dimension $C(N,N/2)$ with $C(. , .)$ the binomial coefficient.

\textbf{\textit{Model 2: Complex NNN coupled $J_1-J_2$}}. The Hamiltonian is given by
\begin{align}\label{Eq14}
        H_2 = J_1  \sum_{\left<i,j\right>}  S_i \cdot S_j
      + (J_2+ i\gamma) \sum_{\left<\left<i,j\right>\right>} S_i \cdot S_j,
\end{align}
where the NNN is now complex. Such complex inter-site coupling can arise from the adiabatic elimination of a dissipative auxiliary mode mediating the NNN interaction. In coupled cavity arrays, complex hopping amplitudes have been engineered via auxiliary lossy resonators~\cite{PhysRevA.81.063807}. In the spin context, a zigzag geometry in which the NNN path passes through a lossy intermediate site would produce an effective complex NNN exchange after tracing out the intermediate degrees of freedom. Non-reciprocal spin interactions, studied in the context of non-reciprocal frustration \cite{PhysRevX.14.011029}, provide a related approach. In contrast to $H_1$, $PT$ symmetry is explicitly broken in $H_2$ for $\gamma \neq 0$.

In $H_2$, the non-Hermiticity is directly introduced into the frustration parameter, and the effective frustration strength is $|J_2+i\gamma|/J_1=\sqrt{J_2^2+\gamma^2}/J_1 \geq J_2/J_1$, which is always at least as large as the bare frustration $J_2/J_1$. Even at $J_2=0$, a purely imaginary NNN coupling of magnitude $\gamma$ introduces frustration where none existed before.

\textbf{\textit{Complex Majumdar-Ghosh locus.}} The Hermitian MG condition (that the exact dimer product state is an eigenstate when $J_2=J_1/2$) generalizes to the complex locus $J_2^2+\gamma^2=(J_1/2)^2$ (analytical prediction), a quarter-circle in the $(J_2, \gamma)$ plane. Along this locus, the dimer product state is expected to satisfy the eigenvalue equation with complex eigenvalues. In summary, $H_1$ introduces non-Hermiticity as a competing energy scale that competes with the exchange, whereas $H_2$ introduces it as an enhancement of the frustration parameter. It is instructive to mention that, in contrast to the TFIM studied above, these models are non-stoquastic, and their ground state has a sign structure; therefore, we used a cRNN for all COMPASS simulations.

\begin{figure*}[ht!]
    \centering
    \subfloat[\centering ]{{\includegraphics[width=0.42\textwidth]{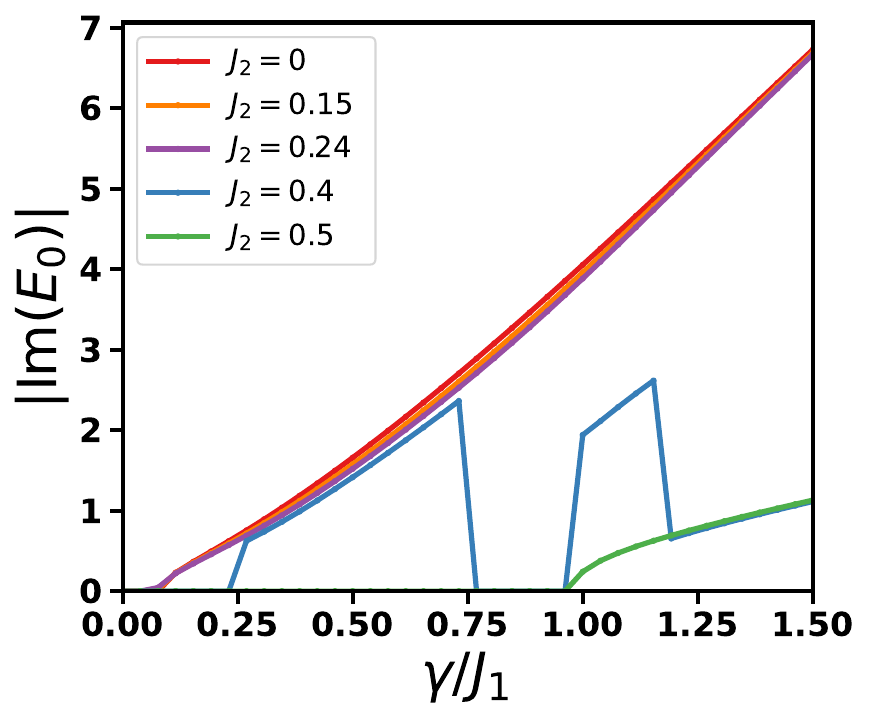} }}
    \subfloat[\centering ]{{\includegraphics[width=0.42\textwidth]{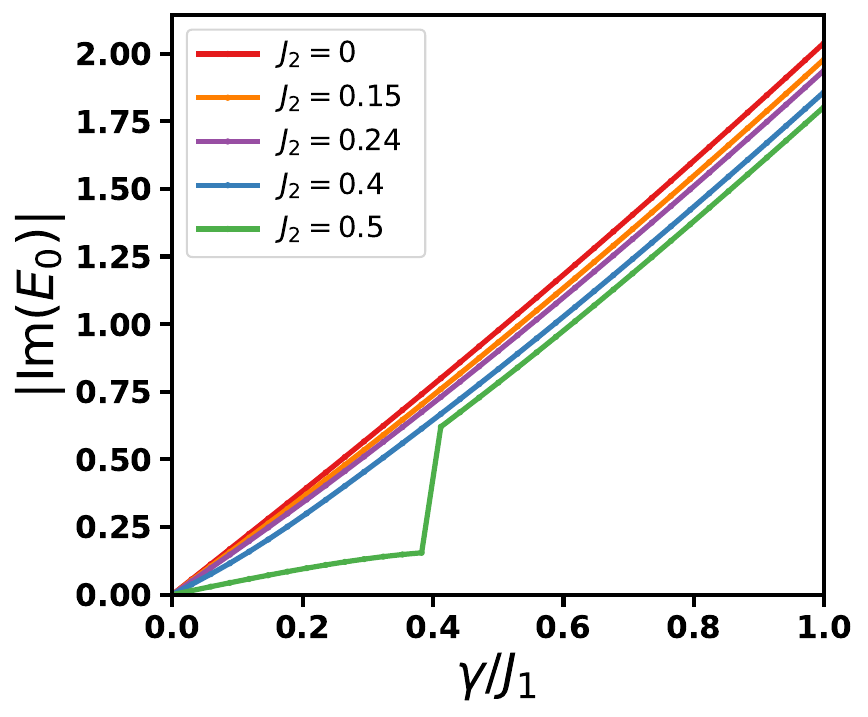} }}
    
    \caption{\textbf{Gap protection of $PT$ symmetry breaking.} We plot the absolute value of the imaginary part of the ground state energy (obtained via ED) in the gapless, $J_2 = 0$ (in red), $0.15$ (orange), and $0.24$ (purple), and the dimerized $J_2 = 0.4$ (blue) phases as well as at the MG point $J_2 = 0.5$ (green). For \textbf{(a)} model $1$ \eqref{Eq13}, we see that the frustration gap provides a quantitative shield against $PT$ symmetry breaking whereas for \textbf{(b)} model $2$ \eqref{Eq14} no such protection is observed. We consider $N=10$, $J_1=1$, and PBCs.}
    \label{F7}
\end{figure*}

\subsection{Non-Hermiticity meets frustration}

\textbf{\textit{Model 1.}} First, we present the results of model $1$, cf. Eq.~\eqref{Eq13}. Fig.~\ref{F6} depicts the phase diagram of the model for various parameters. At $\gamma=0$, $H_1$ reduces to the Hermitian $J_1-J_2$ model. We verify that all eigenstates are purely real [see Fig.~\ref{F6}(a,b)], and the ground-state energy is $E_0=-3N/8$ [see Fig.~\ref{F6}(a,b)] at the MG point with an exact twofold degeneracy ($E_0=E_1$). Indeed, the gap $\Delta=E_1-E_0$ vanishes at this point [see dashed line in Fig.~\ref{F6}(c)]. In addition, the finite-size gap fades near $J_2/J_1\approx 0.28$ for $N=10$, consistent with the Okamoto-Nomura (ON) critical point shifted by finite-size effects [see dotted line in Fig.~\ref{F6}(c)]. We note that, here $\Delta=E_1-E_0$ is computed within the same $S^z=0$ (as we shall discuss later) sector, so it measures  the intra-sector level spacing rather than the conventional singlet-triplet gap. In this convention, the gap decreases as $J_2/J_1$ increases towards the MG point, and the ON transition appears as a gaped-to-gapless crossover, which is the reverse of the standard framing in which the singlet-triplet gap opens at the ON. Both descriptions reflect the same underlying physics: the ON point separates the Luttinger liquid phase from the dimerized phase. In Fig.~\ref{F6}(b), one can clearly see the unbroken-to-broken phase transition that occurs when the eigenvalues move from purely real (white region) to purely imaginary or complex (patched regions) or vice versa. Interestingly, the boundary of this transition bugles outward for large $\gamma$ around $J_2/J_1 \approx 0.5$, where the dimerization gap is the largest, demonstrating a gap-protection mechanism: the finite spectral gap $\Delta$ prevents the ground-state eigenvalue from acquiring an imaginary part up to a critical NH strength $\gamma_c \propto \Delta$. Indeed, this gap protection mechanism is rooted in the structure of $PT$-symmetric spectra: eigenvalues can only acquire an imaginary part by passing through an EP at which two eigenvalues coalesce. Upon reaching such coalescence, the real energies ($E_0$ and $E_1$) are ``equal'' and so the gap $\Delta$ is closed. In order to open this gap, one require a NH perturbation at least of the order of $\Delta$. At a critical $\gamma_c(J_2) \propto \Delta(J_2)$, the $PT$-symmetry is broken and the energies splits into complex conjugate pairs. As result, $\Delta$ act as a ``shield'' to $PT$-symmetry breaking, and the $PT$-breaking threshold tracks the gap $\gamma_c(J_2) \propto \Delta(J_2)$. In addition, in the broken phase, one observes rapidly alternating purple and orange patches, which is a signature of dense level crossings in the complex energy plane. Indeed, when the spectrum is complex, the eigenvalue with the smallest real part changes identity at a dense set of parameter values, producing discontinuous jumps in Im$(E_0)$.

We further observe that the spectral gap reveals well-defined islands coinciding with the $PT$-unbroken region, cf. Fig.~\ref{F6}(c). At the MG point, the gap is exactly zero at $\gamma=0$ (twofold MG degeneracy) but opens immediately for $\gamma>0$ (that is $\Delta(J_2=0.5, \gamma) \propto \gamma$, provided $\gamma \to 0^+$). This observation suggests that non-Hermiticity lifts the MG degeneracy. Indeed, the NH field breaks translational symmetry, distinguishing even- and odd-bond dimer patterns by coupling differently to them, thereby splitting them into two degenerate eigenstates with distinct real eigenvalues. This amounts to the NH perturbation acting as an explicit symmetry-breaking field for the dimer order parameter.

\begin{figure*}[ht!]
    \centering
    \subfloat[\centering ]{{\includegraphics[width=0.33\textwidth]{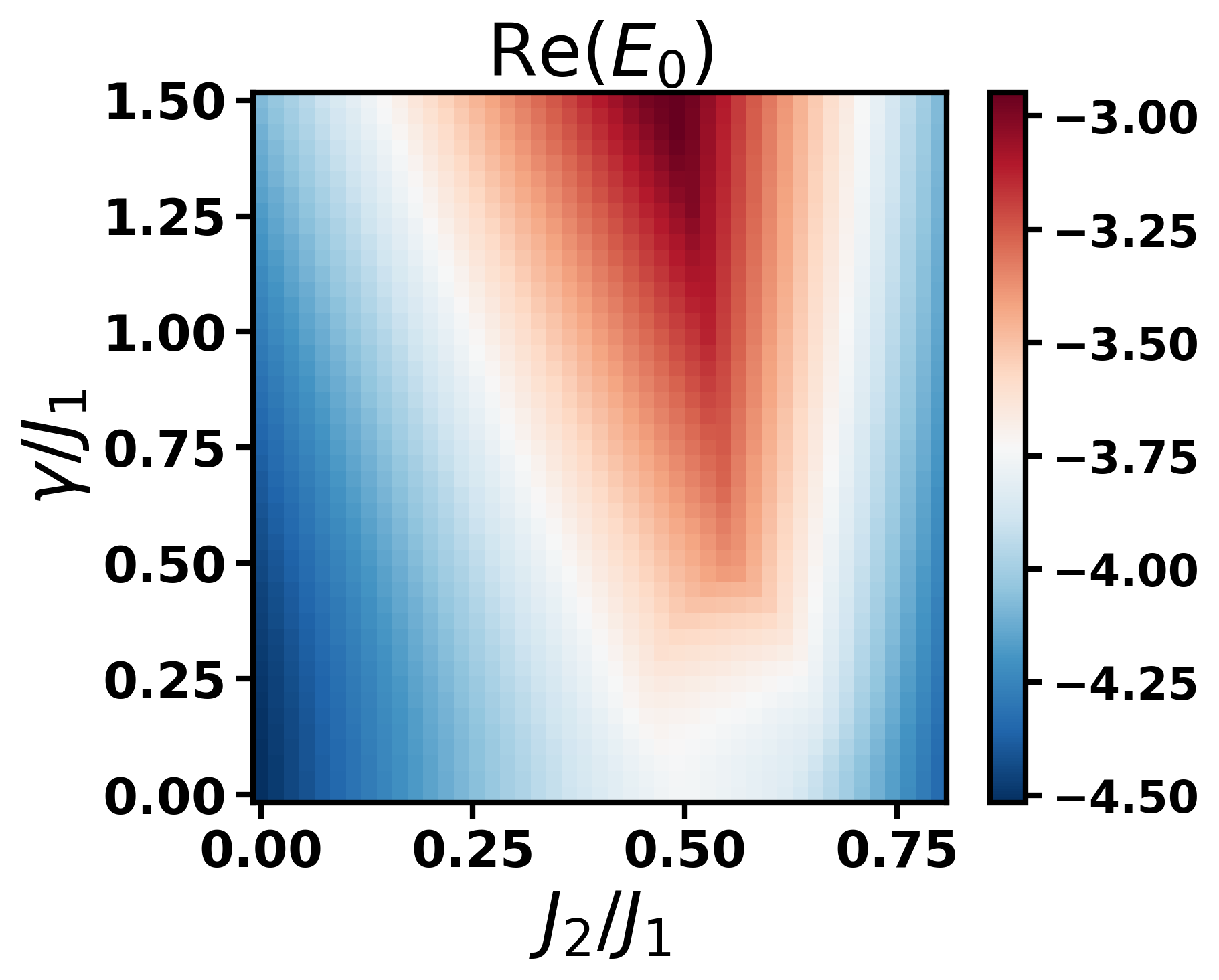} }}
    \subfloat[\centering ]{{\includegraphics[width=0.33\textwidth]{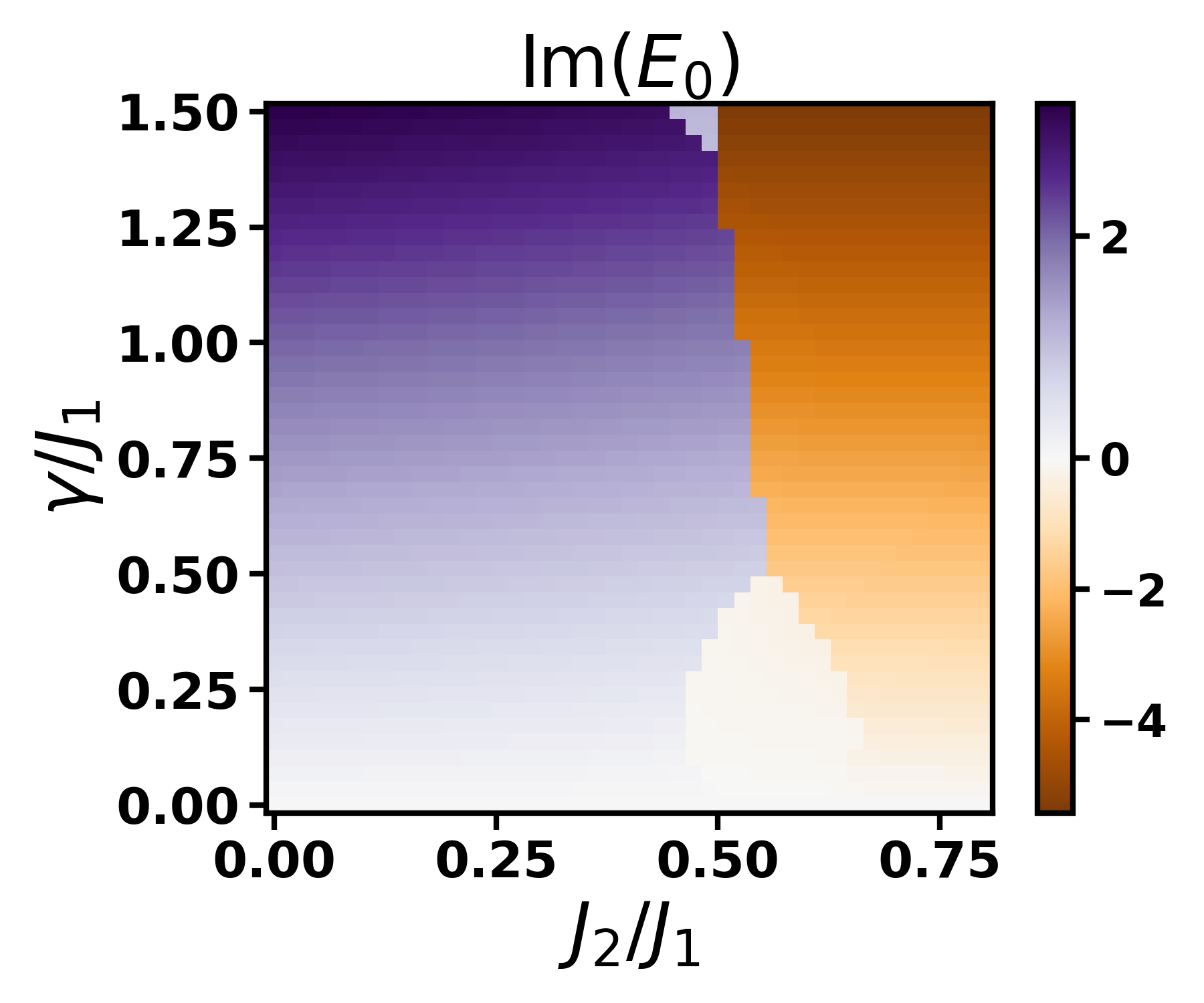} }}
    \subfloat[\centering ]{{\includegraphics[width=0.33\textwidth]{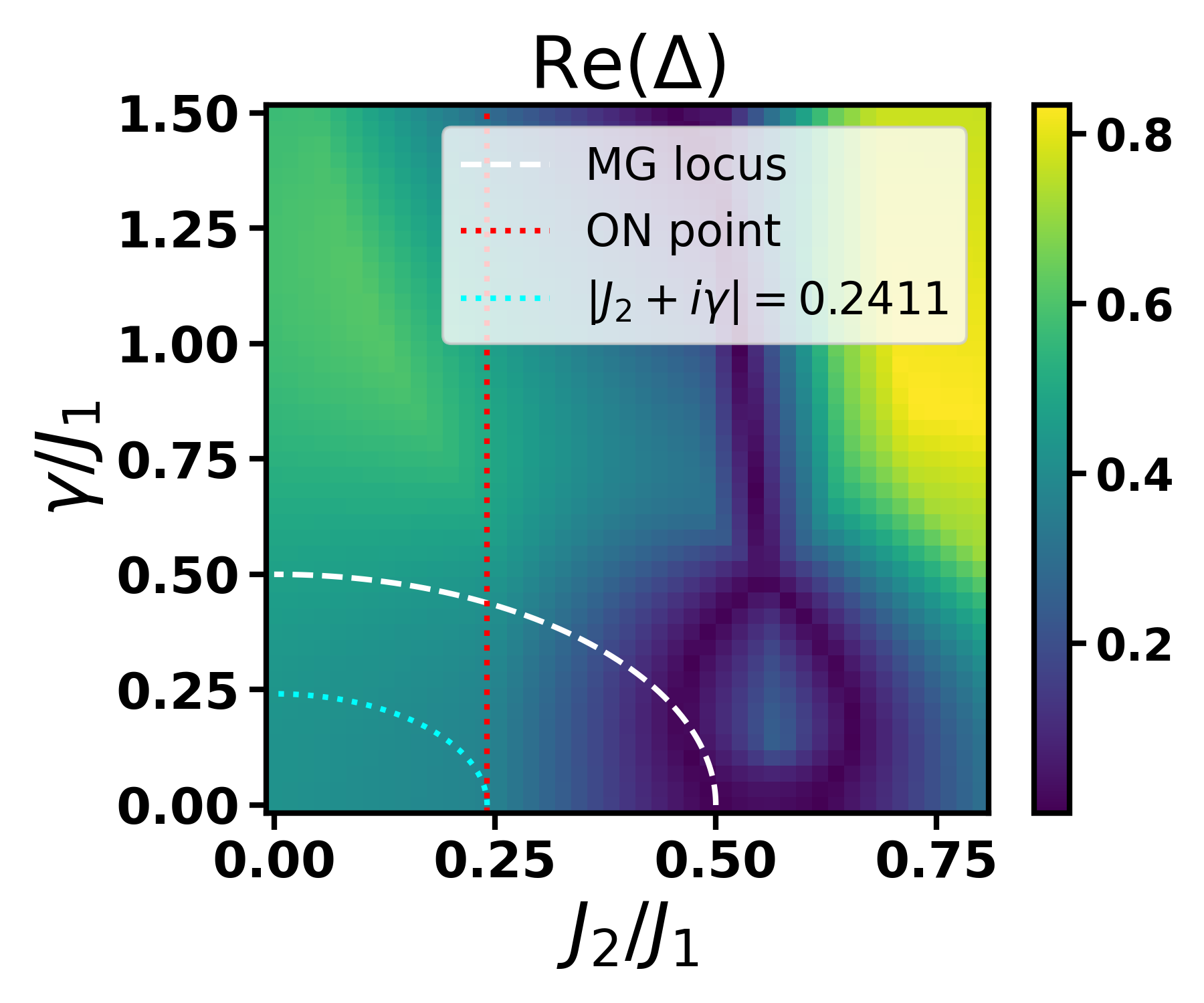} }}
    
    \caption{\textbf{Phase diagram of the NH $J_1-J_2$ with complex NNN coupling}. We show the phase diagram of the Hamiltonian in Eq.~\eqref{Eq14} presenting how the \textbf{(a)} real part, and the \textbf{(b)} imaginary part of the ground state energy $E_0$, as well as the \textbf{(c)} real part of the energy gap $\Delta=E_1-E_0$ changes with non-Hermiticity $\gamma/J_1$ and the frustration parameter $J_2/J_1$. $E_1$ is the first excited state energy obtained from block-diagonalization. The blue (red) line (locus) is the predicted (conventional) ON transition point, while the white dashed line is the predicted MG locus. We considered $N=10$, $J_1=1$ and PBCs. In this case, one observes the emergence of a diabolic ring. Here we used ED.}
    \label{F8}
\end{figure*}

Further analysis of the response of the imaginary part of the energy, shown in Fig.~\ref{F7}(a), reveals that for the gaped phases ($J_2 = 0, 0.15,$ and $0.24$), all the curves collapse onto each other, where the excitation gap is vanishingly small at finite $N$. Without a gap to protect the spectrum, $PT$ symmetry breaks immediately at any $\gamma>0$ and $|\text{Im}(E_0)|$ increases linearly with $\gamma$. The slope is nearly identical for all three because they share the same gapless low-energy structure: the staggered field couples directly to the gapless spin excitation. At $J_2=0.4$, one enters the dimerized phase (blue curve), which shows partial protection, and the imaginary part of the energy remains zero until $\gamma_c\approx0.25$, then jumps. This is the frustration gap at work: the dimerized phase has a finite singlet-triplet gap $\Delta$, and perturbation theory says the imaginary field cannot make eigenvalues complex until $\gamma$ exceeds $\Delta$. The sharp jumps afterward are level crossings, where the ground state switches identity as $\gamma$ pushes different states through each other. At the MG point (green curve), the system has the strongest protection (maximum frustration), and $|\text{Im}(E_0)|$ remains exactly zero until $\gamma_c\approx0.95$. This is because the MG point has the largest gap in the entire dimerized phase; as a result, the exact dimer product state is maximally protected (the protection is nearly four times larger than that at $J_2=0.4$). Overall, the frustration gap provides a quantitative shield against $PT$ symmetry breaking. This establishes a direct connection between gap frustration and NH spectral stability.

\textbf{\textit{Model 2.}} Now, we move on to model $2$ with results shown in Figs.~\ref{F7}(b) and \ref{F8}. In the $J_1-J_2$ model, frustration arises from the competition between the antiferromagnetic NN coupling $J_1$ and the NNN coupling $J_2$. Like model $1$, this model exhibits a region of zero Im$(E_0)$, but it is significantly smaller (along $\gamma=0$) [see Fig.~\ref{F8}(b)], which is consistent with the fact that within this region, the model is still Hermitian, and we can also verify that the energy is $E_0=-3N/8$ [see Fig.~\ref{F8}(a)]. In contrast to $H_1$ where the non-Hermiticity is a local diagonal operator that does not directly modify the exchange couplings creating the gap, in $H_2$, the non-Hermiticity lives inside the NNN coupling itself, and it is then harder for the gap to shield against a perturbation that resides in the same term that creates the gap [see Fig.~\ref{F7}(b)]. The Im$(E_0)$ map is smooth throughout the parameter space, with a well-defined sign structure (uniformly negative for $J_2/J_1<0.5$ , uniformly positive for $J_2/J_1>0.5$). This reflects the absence of dense level crossings: the complex NNN coupling moves the eigenvalues continuously in the complex plane without the competing-energy-scale mechanism that drives rapid level switching in $H_1$.

In Fig.~\ref{F8}(c), one observe a ring of $\Delta=0$ which in contrast to Fig.~\ref{F6}(c) does not represent an exceptional ring (as will be discussed later) but rather a network of level crossings [see Appendix~\ref{A9}]. At $J_2=0.5$, $\gamma=0$, two exactly degenerate ground states (the two dimer coverings) are observed. In this case, the MG degeneracy is preserved because the complex NNN does not break any translational symmetry. When we turn on $\gamma$, the two states acquire complex energies that trace trajectories in the complex plane, while their real parts can ``recross'' at specific ($J_2,\gamma$) points even after the degeneracy is initially lifted. The ring observed in Fig.~\ref{F8}(b,c) is the locus where $\text{Re}(E_1)=\text{Re}(E_0)$, that is, the real parts of the two lowest eigenvalues cross each other. At each crossing point, the states have the same real part but different imaginary parts, which is exactly why Im$(E_0)$ changes sign sharply there, when Re$(E)$ values cross, the identity of the ground state switches to a different eigenstate that has a different Im$(E)$, that is, if state $A$ has Im$(E) < 0$ and state $B$ has Im$(E) > 0$, the sign flips at the crossing. Overall, the two MG descendants respond differently to the complex coupling. Near $J_2/J_1=0.5$ the perturbation splits them along Im$(E)$ first. However, as we move in $J_2$ away from $0.5$, the real frustration also reshuffles the Re$(E)$ ordering. The crossing occurs on a close curve because the perturbation has a smooth structure. Later, for $\gamma>0.5$, one observes a line of $\Delta=0$, while this could reflect crossings of the two MG descendants continuing beyond the ring, it may also signal that higher excited states are pushed down into the low-energy manifold at larger $\gamma$. Distinguishing these two scenarios would require tracking eigenvalues explicitly, which we leave for future work. We argue that the complex NN coupling creates a topologically nontrivial network of real-energy level crossings (diabolic points) emanating from the Hermitian MG degeneracy, which is distinct from EPs (as we will discus later). The region enclosed by the innermost crossing ring coincides with the ``accidentally real'' spectrum pocket, establishing a connection between the MG dimer degeneracy and the NH spectral stability.

We naively predict the ON (cyan quarter-circle) and the MG (white quarter-circle) by setting $|J_2 +i\gamma| \approx 0.24$ and $0.5$, respectively. In doing so, we effectively assume that we can consider $|J_2 +i\gamma|$ as an effective coupling controlling the frustration, or in other words, we assume the real ($J_2$) and complex ($\gamma$) NNN coupling are interchangeable. This hypothesis is motivated by the fact that we assumed that the competition between the NN and NNN bonds is controlled by the magnitude of the coupling, since in the mean-field or variational sense, what matters is how strongly the NNN bonds competes with NN bonds. Also, one can view $\gamma \neq0$ as a rotation of $J_2$ in the complex coupling plane, and in general in analytically continued models, phase boundaries often only depends on the magnitude of the coupling.

If considering $|J_2 +i\gamma|$ as an effective coupling is a correct assumption, one should observe a phase transition along the entire cyan quarter-circle. However, Fig.~\ref{F8}(c) shows that there is no feature along this curve: Indeed, regardless of the values of $\gamma$, Re$(\Delta)$ evolves smoothly (no opening or closing of the gap), and the ON transition does not extend into the complex plane along the constant contour $|J_2+i\gamma|$. This directly disproves the universal effective frustration hypothesis, which we assumed at the beginning of this paragraph. Likewise, for the MG point, we see that the exact degeneracy does not persist along the entire white quarter-circle. However, the ring of level crossings cuts through the MG locus at roughly $\gamma \approx 0.4$. This means that as one moves along the MG circle away from the Hermitian point, one hits a level crossing where the ground-state identity changes---the MG descendants are no longer the lowest states. The MG solution does not survive the complexification along this circle. 

Surprisingly, the actual phase structure is shaped by the ring, not by the predicted quarter-circles. The ring encloses a region, where the two MG descendants remain in the lowest-energy states, and the gap between them is the relevant gap. Outside the ring, other states or the same lowest-energy states may have crossed. This ring boundary is determined by the detailed level structure that is how each eigenstate responds individually to real versus imaginary coupling. Overall, the cyan quarter-circle were supposed to predict the gaped-to-gapless phase transition, and the white one the closed gap. Neither of the predictions worked. The actual gap structure follows the ring, which has a completely different topology from either quarter-circle. This proves that complexifying $J_2$ creates qualitatively new physics that cannot be reduced to an effectively real frustration parameter.

\subsection{Emergence of novel spectral topology in non-Hermitian frustrated systems: The diabolic ring}

In this subsection, we further investigate the effect of non-Hermiticity on frustration and summarize our results. We only focus on model $2$, cf. Eq.~\eqref{Eq14}, since the competition is more visible, and we identify a closed locus of diabolic level crossings in the parameter space---a diabolic ring---that supports eigenvalue braiding and has no Hermitian analog. Unlike exceptional rings, where eigenvectors coalesce, diabolic rings preserve eigenvector distinguishability, while permitting real-energy crossings between states with distinct imaginary parts.

First, in Hermitian quantum mechanics, the von Neumann-Wigner theorem states that energy levels repel~\cite{vonNeumann1993} meaning they generically avoid crossing. This is why the Hermitian MG degeneracy is special: it requires fine-tuning (exactly $J_2=J_1/2$) and is protected by the symmetry. In NH systems, the levels can attract each other. Two eigenvalues that have different imaginary parts are free to cross in their real parts. Indeed, there is no repulsion between Re$(E)$ values when the Im$(E)$ values differ. The ring we observe is a closed curve of such crossings: two states approach in Re$(E)$, pass through each other, and separate again. A closed ring of crossings between states in the same sector as observe here, has no Hermitian analog. The ring topology (closed loop rather than isolated points) arises because crossings in the $(J_2, \gamma)$ plane are codimension-1 curves (one condition Re$(E_1)=$ Re$(E_0)$), not codimension-2 points as in the Hermitian case (the two MG dimer states carry different lattice momenta $k=0$ and $k=\pi$, which forces their off-diagonal matrix element to vanish by translation symmetry, reducing the crossing from codimension-3 to codimension-2. The MG degeneracy is therefore 
already a fine-tuned event in the 1D parameter space $J_2/J_1$.).

Furthermore, the complex NNN coupling preserves $SU(2)$ symmetry ($S\cdot S$ is a scalar), but breaks time reversal $T$. In a Hermitian system, time-reversal symmetry constrains the eigenvalues as $\{E(k)\} = \{E(-k)\}$. Breaking $T$ lifts this constraint. However, the two MG dimer states are related by on-site translation, not by $T$. Therefore, when $T$ is broken by $i\gamma$, the two dimers acquire imaginary energies at different rates. One gets pushed down in Im$(E)$, and the other up. This is why Im$(E_0)$ changes sign at the ring edge. A real $J_2$ cannot do this since it preserves $T$, and both dimers always have the same (real) energy evolution, this is a channel-selective symmetry breaking.

Moreover, as we move around a closed loop in $(J_2, \gamma)$ space that encircles the ring, eigenvalues can braid, and $E_0$ and $E_1$ exchange identity after one circuit is completed. This is a spectral manifestation of NH topology, which we refer to as spectral flow topology. This ring is the branch cut of this braiding: crossing it switches which eigenvalue one can call ``ground state''. The white pocket in Im$(E_0)$ is the region where the spectrum remains accidentally real and the ground state identity is globally consistent: throughout this pocket, the same MG-descended eigenstate has the lowest real part and zero imaginary part. Outside the pocket, the ring boundary has been crossed and the ground state label has switched to a different eigenstate, making a global consistent labeling impossible. This braid has no Hermitian analog. In Hermitian systems, states can always be labeled by energy ordering, and the labeling is globally consistent. In NH systems, the complex nature of eigenvalues means that the ordering Re$(E_0)<$ Re$(E_1)$ can be locally defined but globally inconsistent, which is exactly what occurs at the ring.

Finally, in the Hermitian limit, $J_2$ enters the Hamiltonian through $J_2S\cdot S$, and all physical properties are analytic functions of this single real parameter. The naive extension will be to replace $J_2\to|J_2+i\gamma|$. This fails because the Hamiltonian is $(J_2+i\gamma)S\cdot S$, not $|J_2+i\gamma|S\cdot S$. In other words, the phase of the complex coupling arg($J_2+i\gamma$) is important, not just its modulus. We argue that the phase determines how much of the coupling goes into Re($E$) and Im($E$), which eigenstates are pushed up and which ones are pushed down in the complex plane, and where the level crossings occur. Therefore, two points with the same $|J_2+i\gamma|$ but different arg($J_2+i\gamma$) see genuinely different Hamiltonians that are not related by any symmetry transformation. We stress that the complex NNN couplings create a NH spectral topology, a ring of diabolic crossings with eigenvalue braiding, that has no Hermitian counterpart. This topology is controlled by the phase of the complex coupling rather than its modulus, establishing that the NH frustration is fundamentally richer than its Hermitian analog extended to complex parameters.

\begin{figure*}[ht!]
    \centering
    \subfloat[\centering ]{{\includegraphics[width=0.49\textwidth]{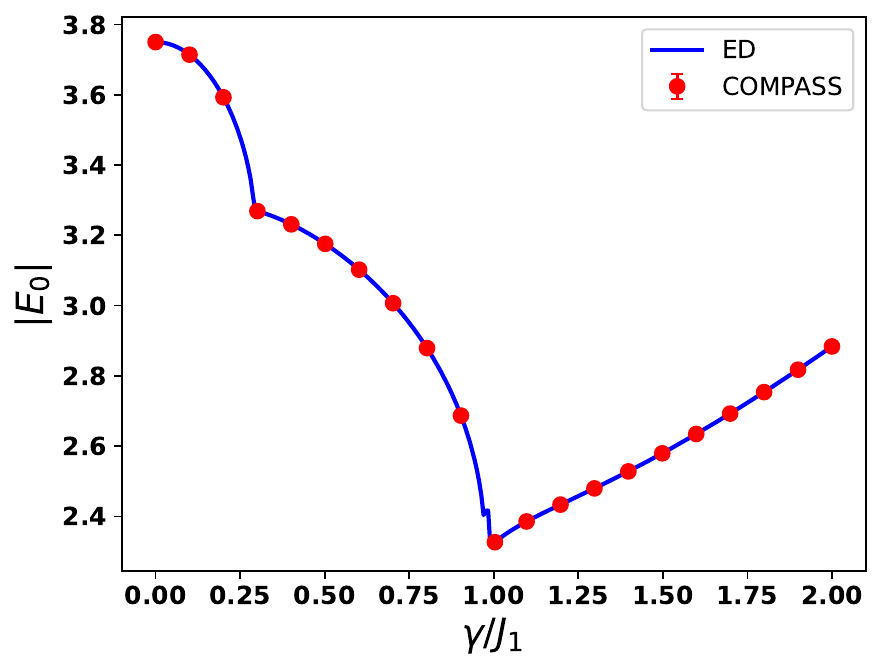} }}
    \subfloat[\centering ]{{\includegraphics[width=0.49\textwidth]{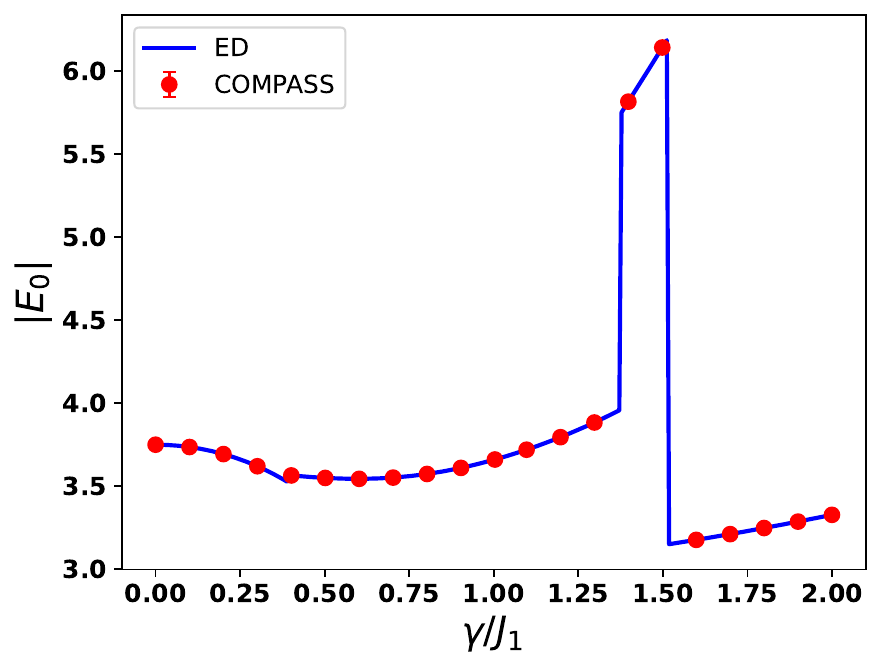} }}
    
    \caption{\textbf{ED vs COMPASS}. We compare ED (blue) and COMPASS (red) for \textbf{(a)} model $1$, cf. Eq.~\eqref{Eq13}, and \textbf{(b)} model $2$, cf. Eq.~\eqref{Eq14}. We consider $N=10$, and $J_2/J_1=0.5$. The COMPASS algorithm is able to consistently reproduce ED.}
    \label{F9}
\end{figure*}

\subsection{Emergence of exceptional points}

For model $1$, cf. Eq.~\eqref{Eq13}, we stress that along the MG point line, the gap closes and opens at three different points (around $\gamma \approx 0.1, 0.4,$ and $1$) and that these transition points are exceptional (EPs) as further confirmed in Appendix~\ref{A9}. As such, the contour of the islands separating the gapless and gaped regions corresponds to an exceptional line (juxtaposition of several EPs) as also confirmed by Fig.~\ref{F6}(c)---as it is now well established that the unbroken-to-broken phase transition generally passes through an EP~\cite{wah2025,RevModPhys.93.015005}. However, for model $2$ \eqref{Eq14} no EPs were actually detected. Although we observe  a peak in the maximum overlap (at $\approx 0.89$) trending towards $1$ when crossing the ring, this maximum overlap does not yield the coalescence of the eigenstates [see Appendix~\ref{A9}]; as mentioned above, we attribute this to a diabolic ring.

\subsection{COMPASS for frustrated systems}

Now that we have established some background on our frustrated models, in this subsection we show that the COMPASS framework developed in this work can also recover the ground-state properties of both models. Fig.~\ref{F9} depicts the magnitude of the ground state energy obtained via ED (blue) and COMPASS (red). For model $1$ [see Fig.~\ref{F9}(a)], the COMPASS method finds the Hermitian ($\gamma=0$) energy ($\approx3.75$), which corresponds to the exact MG point and decreases down to a minimum around $\gamma \approx 1$ due to the fact the staggered magnetic field weakens the dimer bonds via second-order perturbation theory. After the breaking of the $PT$ symmetry, however, the ground state identity switches and $|E_0|$ starts increasing as Im($E_0$) grows. The smooth curve with sharp features reflects the gradual erosion of the MG state. Conversely, in model $2$ [see Fig.~\ref{F9}(b)], one observe a kink at $\gamma \approx[1.4, 1.5]$, which corresponds to a level crossing (the same one forming the boundary of the diabolic ring). Indeed, for $\gamma < 1.4$ the ground state descends from the MG doublet that has moderate $|E_0|$ with both real and imaginary part growing. At $\gamma \approx 1.4$, a completely different eigenstate---one that was previously an excited state with large $|E|$---crosses below in Re($E$). This new ground state has much smaller $|\textrm{Re}(E_0)|$ but also different Im($E_0$), producing the sharp drop from  $\approx 6$ to $\approx 3.2$. The spike just before the drop is the approaching state's energy being large right before the crossing. This is essentially a 1D slice through the diabolic ring, and the sharpness of the kink confirms that it is a true level crossing (codimension-$1$), not a smooth crossover. Interestingly, the NQS is able to capture this physics, which further confirms the robustness of the COMPASS method. We limited ourselves to a small system size, where results can be benchmarked against ED, and while we can, in principle, simulate larger system sizes, this will require either an exactly solvable model (which is not the case for our models) or the development of an appropriate density matrix renormalization group algorithm for non-Hermitian frustrated systems, which we deliberately leave for future work. It is worth mentioning that the provided framework remains valid in the Hermitian regime ($\gamma=0$) as consistently shown by our analysis.

\section{Conclusion}\label{S5}

In this work, we introduce COMPASS---a biorthogonal adaptive neural quantum state framework---that establishes a principle and versatile approach for studying NH many-body-systems. Our results reveal critical insights into the role of the choice in ansatz for NH simulations. For $PT$-symmetric systems, we demonstrate that unconstrained ans\"atze can spontaneously break the $PT$ symmetry during optimization,even within the nominally unbroken phase, yielding spurious imaginary energies. Real-valued ans\"atze naturally enforce the correct physical manifold, preserving the spectral reality expected in the unbroken regime. Conversely, for generic NH Hamiltonians, complex ans\"atze become indispensable for accurately capturing complex ground-state properties. These findings establish that physically informed ansatz selection is not merely a technical detail but a foundational requirement for reliable NH simulations, and that physical fidelity requires architecture and symmetry-informed modeling, rather than maximal expressivity. Interestingly, our method enables the simulation of 1D NH systems up to $N=200$ and 2D NH systems up to $N=100$ spins, which at the best of our knowledge, is the largest system size ever reported in both dimensions.

Beyond the methodological advances, our work opens a new frontier at the intersection of non-Hermiticity and frustrated magnetism. By extending COMPASS to two physically motivated NH extensions of the paradigmatic $J_1-J_2$ chain, we uncovered previously unexplored phenomena that have no Hermitian analog. We showed that gap frustration acts as a quantitative shield against NH spectral instability, with the frustration gap setting a critical threshold for $PT$ symmetry breaking. Moreover, complexifying the frustration coupling itself engenders a topologically nontrivial network of diabolic level crossings, namely, a diabolic ring,controlled entirely by the phase of the complex coupling. These results demonstrate that frustrated NH systems host a rich and largely uncharted landscape of emergent spectral topology.

Taken together, our framework enables the direct study of 1D and 2D NH many-body systems without reliance on Hermitian embeddings, adiabatic continuations, or symmetry simplifications as reported in all previous works. COMPASS positions NQS as a robust and physically informed tool for exploring the full complexity of NH quantum matter regardless of the dimensionality (1D or 2D Hamiltonians) and the stoquasticity, or in the presence of frustration. Looking forward, this approach opens several directions: extending COMPASS to time-dependent NH dynamics, exploring exceptional points and topological phases, and integrating adaptive NQS with experimental quantum platforms for real-time characterization of open-system phenomena. In addition, it could be interesting to further study biorthogonal variance as a clean numerical signature of $PT$-phase transition as hypothesized in this work [see Appendix~\ref{A3} for the mathematical motivation]. More broadly, the synergy between adaptive variational methods and biorthogonal quantum geometry laid out here provides a foundation for systematically uncovering new NH phenomena that remain inaccessible to conventional numerical approaches.

\textit{Note added:} During the finalization of this study, a related study~\cite{atif} on the importance of choosing the appropriate antsatz for dissipative quantum systems appeared. The work extensively presents complex RNNs as natural ans\"atze for dissipative system with complex parameters. They showed that for generic NH system, an cRNN is crucial for correctly representing the quantum state. this conclusion is in complete agreement with our results.

\section{Acknowledgments}

L.W. and F.K.K. acknowledge funding from the Max Planck Society Lise Meitner Excellence Program~\mbox{2.0}. F.K.K. also acknowledges funding from the European Union's ERC Starting Grant ``NTopQuant'' (101116680). M.H. acknowledges support from the Natural Sciences and Engineering Research Council of Canada (NSERC) and the Digital Research Alliance of Canada. Views and opinions expressed are, however, those of the authors only and do not necessarily reflect those of the European Union or the European Research Council (ERC). Neither the European Union nor the granting authority can be held responsible for them.

\section{Data Availability}

The code that support the findings of this article are
openly available at \cite{GitHubRepo}; embargo periods may apply.


\appendix

\section{Biorthogonal variational Monte Carlo}\label{A1}

\begin{figure*}[ht!]
    \centering
    \subfloat[\centering ]{{\includegraphics[width=0.5\textwidth]{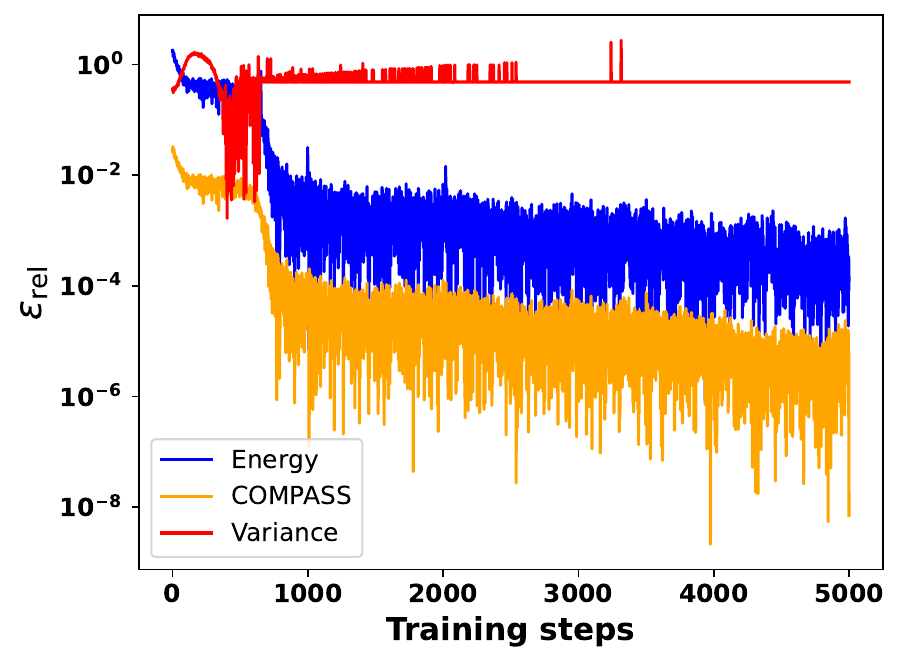} }}
    \subfloat[\centering ]{{\includegraphics[width=0.5\textwidth]{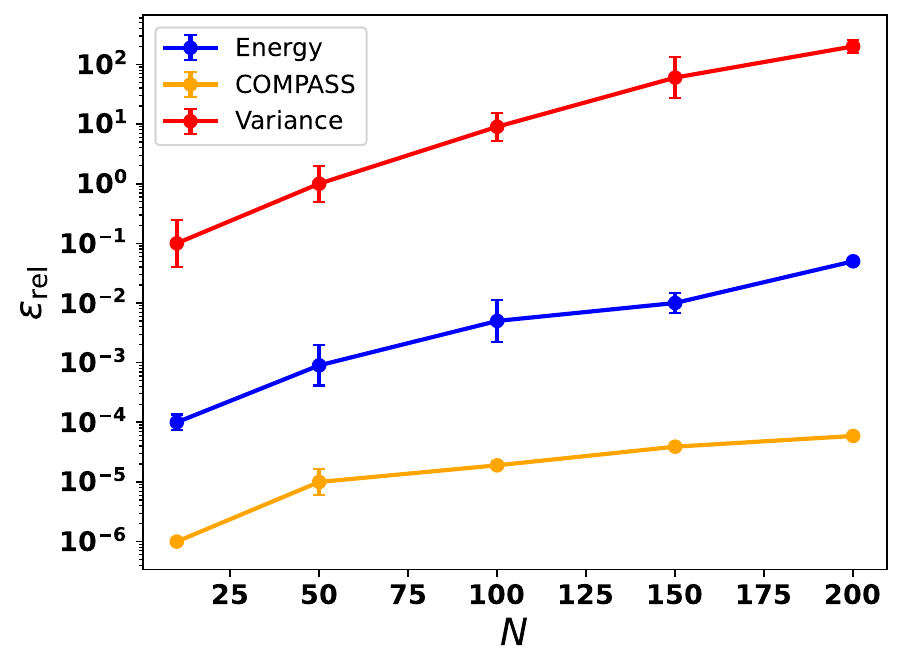} }}
    
    \caption{\textbf{Optimization schemes}. We compare the relative error $\varepsilon_{\text{rel}}$ achieved with pure energy minimization (blue), variance refinement (red), and the complementary optimization (yellow). We plot $\varepsilon_{\text{rel}}$ as a function of \textbf{(a)} the training steps for $N=10$, and \textbf{(b)}~the system size $N$. OBCs are considered. The complementary approach clearly outperforms both energy and variance minimization. For $N=10$ we compare with ED, and for $N=50, 100, 150, \text{and} \,\, 200$ we compare with SE. The model considered is the 1D $PT$-symmetric TFIM in Eq.~\eqref{Eq1}.}
    \label{FA5}
\end{figure*}

In general for Hermitian systems, VMC is based on the Rayleigh-Ritz variational principle, which states that for a Hermitian operator $H$, the ground-state energy is
\begin{equation}\label{E1}
E_0 = \min_{\Psi \neq 0}
\frac{\langle \Psi | H | \Psi \rangle}{\langle \Psi | \Psi \rangle},
\end{equation}
and for any normalized trial state $|\psi\rangle$,
\begin{equation} \label{E2}
\langle \Psi | H | \Psi \rangle \ge E_0.
\end{equation}
Eq.~\eqref{E1} is called a pure estimator, since the same wavefunction is used to sandwich $H$.

For a non-Hermitian system, one instead has a mixed estimator
\begin{equation}\label{E3}
E_0 = 
\frac{\langle \Psi^L_0 | H | \Psi^R_0 \rangle}{\langle \Psi^L_0 | \Psi^R_0 \rangle},
\end{equation}
where for Hermitian systems, $\langle\Psi^L_0|$, and $|\Psi^R_0\rangle$ always satisfy the Hermitian conjugate $\langle\Psi^L_0|=(|\Psi^R_0\rangle)^{\dagger}$. In this case, sampling is done over $\rho(\bm{x})=|\Psi_0(\bm{x})|^2$. However, for generic non-Hermitian systems $\langle\Psi^L_0|\neq(|\Psi^R_0\rangle)^{\dagger}$, and spin configurations are sampled from 
\begin{align}
    \rho(\bm{x})=\frac{|\Psi^L_0\Psi^R_0|}{\sum_{\bm{x}}|\Psi^L_0\Psi^R_0|}.
\end{align}

This procedure is refer to as biorthogonal VMC (bVMC). It has been shown that one may also only use the right or left eigenstates to draw these samples \cite{solinas2025biorthogonal,wah2025}. It is instructive to note that for non-Hermitian system, if one forms a matrix with all right eigenstates as columns and then takes its inverse, the left eigenstates can be obtained as the rows of that inverse matrix, which simplifies the optimization process. While one cannot use this relation here, we can always take advantage of the symmetries present in the system to verify if one can optimize only one state, e.g., the right state, and apply the symmetry to recover the other state, e.g., the left state.

\section{Loss functions}\label{A2}
The optimization process is performed using the energy-based and the variance-based loss functions. First, we construct the energy loss function $\mathcal{L}_e$ such that
\begin{align}
    \mathcal{L}_e &= \mathcal{L}_e^L + \mathcal{L}_e^R,
\end{align}
where the left and right loss $\mathcal{L}_e^L$, $\mathcal{L}_e^R$ are
\begin{align}
    \mathcal{L}_e^R=\frac{\langle \Psi^R|H|\Psi^R\rangle}{\langle \Psi^R|\Psi^R\rangle}, \qquad \mathcal{L}_e^L=\frac{\langle \Psi^L|H^{\dagger}|\Psi^L\rangle}{\langle \Psi^L|\Psi^L\rangle}.
\end{align}

Likewise, the biorthogonal variance-based loss function is constructed as \cite{solinas2025biorthogonal}
\begin{align}
    \mathcal{L}_v &= \mathcal{L}_v^L + \mathcal{L}_v^R,
\end{align}
where
\begin{align}\mathcal{L}_e^R=\frac{\langle \Psi^R|\sigma^2_R|\Psi^R\rangle}{\langle \Psi^R|\Psi^R\rangle}, \,\ \mathcal{L}_e^L=\frac{\langle \Psi^L|\sigma^2_L|\Psi^L\rangle}{\langle \Psi^L|\Psi^L\rangle},
\end{align}
and
\begin{align}\sigma^2_R(\varepsilon)=(H^{\dagger}-\varepsilon^*)(H-\varepsilon),\, \sigma^2_L(\varepsilon)=(H-\varepsilon)(H^{\dagger}-\varepsilon^*),
\end{align}
where $\varepsilon=\langle\Psi^L|H|\Psi^R\rangle/\langle\Psi^L|\Psi^R\rangle$ is incorporated self-consistently via bVMC, and the two variances vanish when the pair of states $|\Psi^R\rangle$, and $\langle \Psi^L|$ are a pair of biorthogonal eigenstates. For both losses, the local energies are treated as constants (stop-gradient) during backpropagation.

\begin{figure*}
    \centering
    \includegraphics[width=0.9\textwidth]{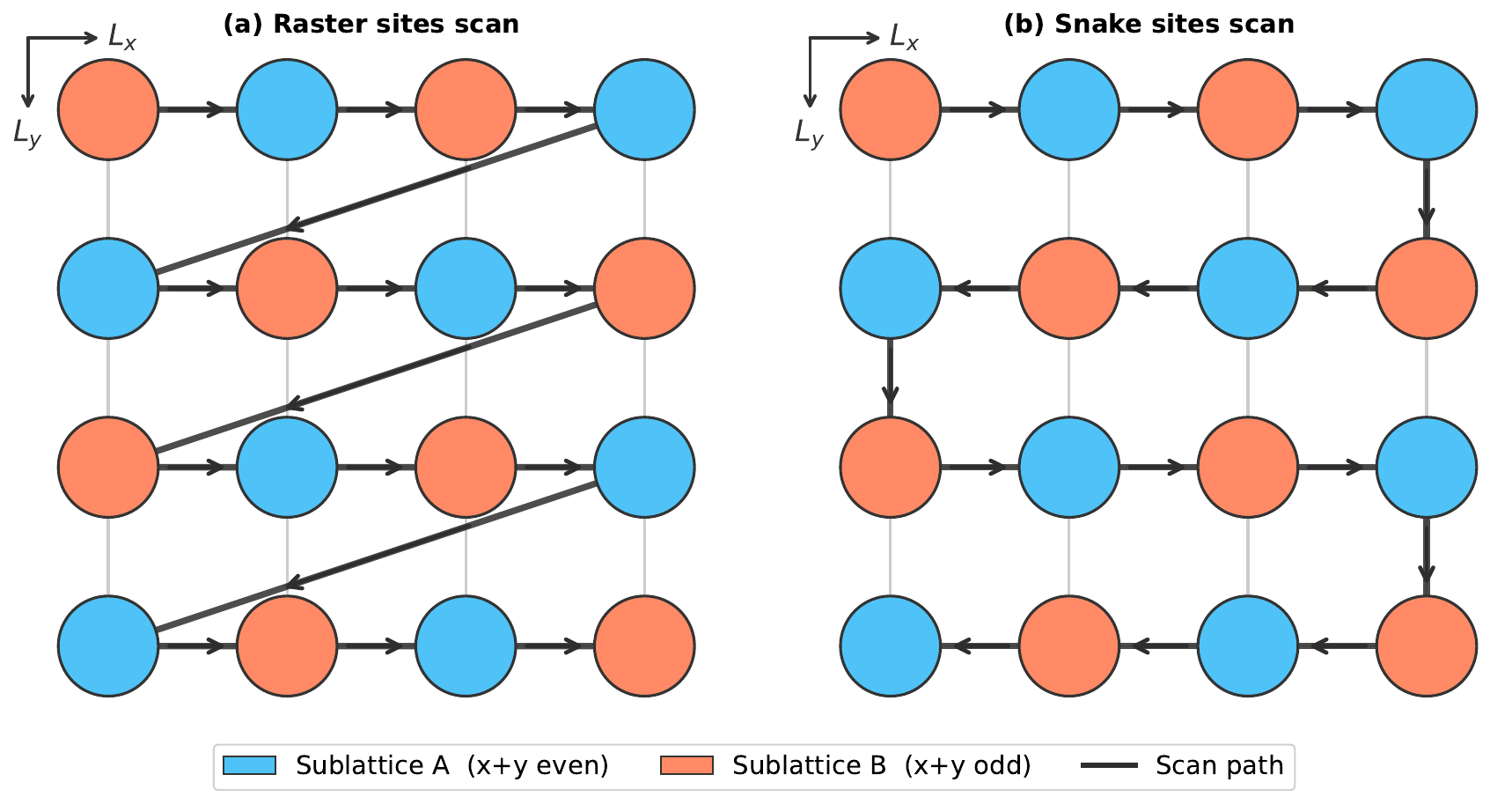}
    \caption{\textbf{Sites scanning techniques}. We illustrate the \textbf{(a)} raster scanning and the \textbf{(b)} snake scanning for the 2D $PT$-symmetric TFIM on a $4 \times 4$ square lattice. The blue circles indicate sites of sublattice $A$ (even), and the orange circles indicate sites of sublattice $B$ (odd). The raster breaks the staggered structure of the model and introduces nonphysical long-range interactions, whereas the snake scan preserves the sublattice symmetry of the model and only accounts for nearest-neighbor interactions. }
    \label{FA6}
\end{figure*}

\section{Variance vs. energy vs. complementary optimization}\label{A5}
Here we compare our optimizations schemes. First, we simulate the ground state of the 1D $PT$-symmetric TFIM in Eq.~\eqref{Eq1} using pure variance minimization. For $N=10$, we observe the convergence to a pair of eigenstates that are not the ground state, resulting in a large relative error in the energy [see Fig.~\ref{FA5}(a)]. This behavior persists as the system size increases [see Fig.~\ref{FA5}(b)], which justifies the need for a warm-start strategy to ensure the NQS converges to the ground-state eigenpair. We also perform pure energy minimization, which performs significantly better than the variance minimization for small system sizes [see Fig.~\ref{FA5}(a)]. However, the relative error remains relatively large and increases linearly with system size [see Fig.~\ref{FA5}(b)]. Finally, we apply our complementary approach, where the error grows much more slowly and remains relatively small as the system size increases~[see Fig.~\ref{FA5}].

\section{Sites scanning and early stopping}\label{A6}
\subsection{Sites scanning}
In Sec.~\ref{S3}, we mention that applying a 1D RNN to 2D lattice model necessitates choosing a scanning path, which can significantly affect the accuracy of the resulting many-body wavefunction. Such inaccuracy arises because certain scanning paths may introduce nonphysical sequential correlations or skip relevant spatial interactions between sites. 

In the present model, however, the snake scan provides a systematic advantage over the raster scan that goes beyond this general consideration. The 2D $PT$-symmetric TFIM studied here features a staggered checkerboard sublattice structure, with sublattice A (blue) and B (orange) alternating on every bond. Under raster scanning, the A/B pattern along each row is regular, but the parity is preserved at row boundaries, meaning two consecutive sites in the sequence (the last site of row $r$ and the first site of row $r+1$) belong to the same sublattice [see Fig.~\ref{FA6}(a)]. 

Moreover, the step across a row boundary connects the far end of one row to the far end of the next row in the opposite direction [see Fig.~\ref{FA6}(a)]. These sites are not physical nearest-neighbor---they are $L_y -1$ lattice spacings apart and share no bond in the Hamiltonian---yet the RNN is forced to perform an autoregressive prediction step between them as if they where sequential. 

The snake scan eliminates this artifact by reversing the scanning direction on alternating rows. As a result, every sequential step across a row boundary always connects sites of opposite sublattices [see Fig.~\ref{FA6}(b)], preserving a perfectly alternating A/B pattern throughout the entire sequence. This alignment between the autoregressive ordering and the physical symmetry of the NH field facilitates learning the correct conditional structure of the ansatz, which is  empirically reflected in a lower variational energy compared to the raster scanning.

\subsection{Early stopping}
\begin{figure}
    \centering
    \includegraphics[width=0.48\textwidth]{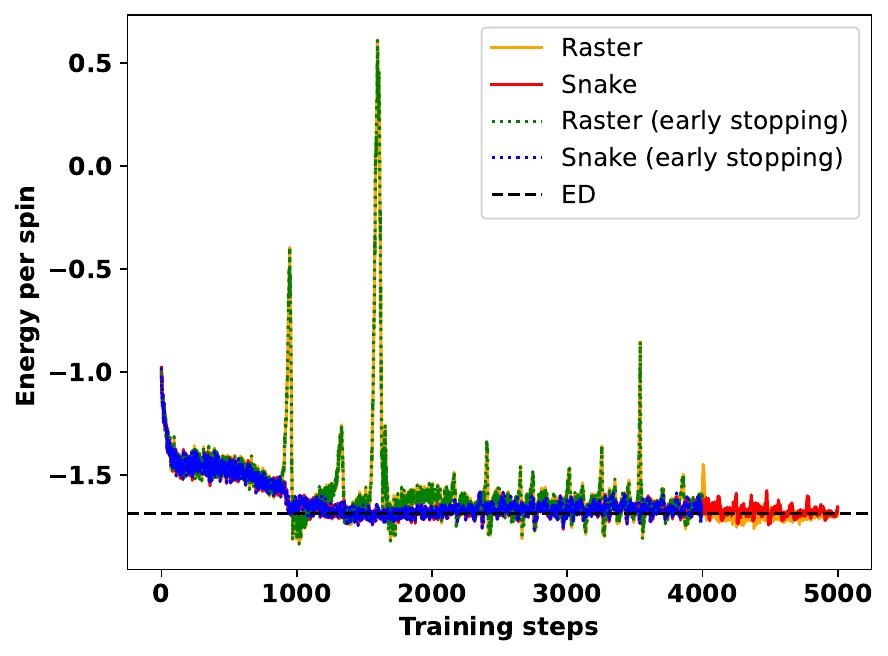}
    \caption{\textbf{Early stopping}. We implement early stopping for the raster (dashed green), and the snake (dashed blue) scan for $L_x \times L_y =3 \times 3$, $\eta=1.6$, and $\xi=\eta/10$. The optimization terminates after approximately $4000$ training steps, and we can indeed see that after those steps, no meaningful improvement is observed in the ground state energy per spin as shown in orange and red for the raster and snake scans, respectively. OBCs are considered. }
    \label{FA61}
\end{figure}

In machine learning and optimization, early stopping \cite{prechelt2002early,PRECHELT1998761} is a fundamental regularization technique designed to prevent overfitting while reducing computational cost. The core idea is to monitor the model's performance on a validation set during training and halt the optimization process when performance ceases to improve. This approach effectively balances underfitting and overfitting by identifying the point at which the model has learned the underlying patterns in the training data without beginning to memorize noise. Early stopping not only enhances the model generalization, but also provides practical benefits by reducing training time and thus, computational cost as depicted in Fig.~\ref{FA61}. 

In our context, we implement early stopping based on energy convergence. Specifically, the optimization is terminated when the change in energy $\Delta E=|E_i-E_{i-1}|< 10^{-5}$ (where $i$ indexes the training step) for a consecutive number of steps $N_{\text{pat}} =500$. In other words, the algorithm ``stops early'' if no further improvement in the energy is observed. Here, $N_{\text{pat}}$ is referred to as the patience parameter.

\section{Fidelity of the eigenpair}\label{A7}

\begin{figure}
    \centering
    \includegraphics[width=0.48\textwidth]{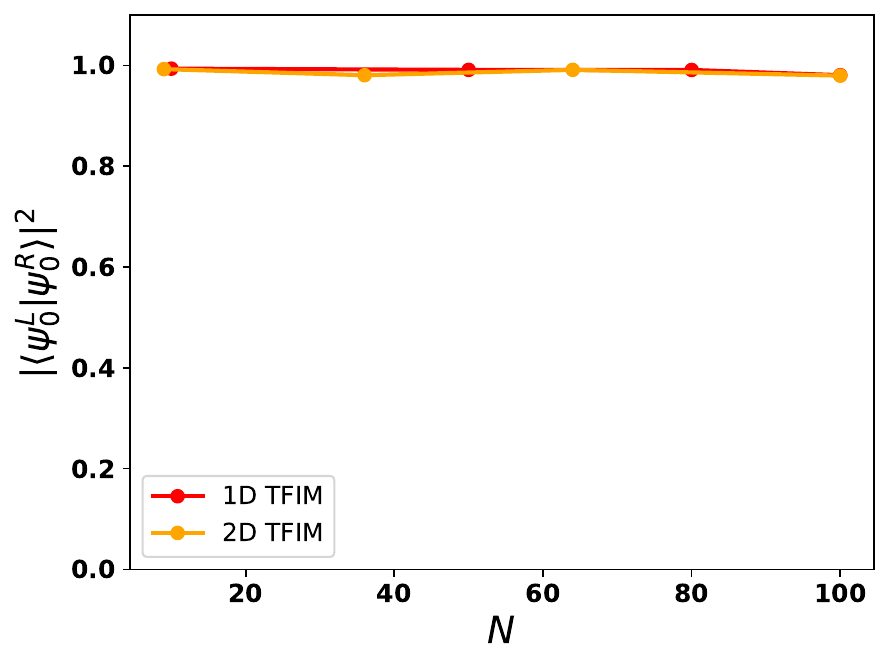}
    \caption{\textbf{Fidelity}. We plot the fidelity for the 1D (red) and 2D (yellow) $PT$-symmetric TFIM for different system size. One observe a high fidelity in both case. We consider $\eta=1.6$, $\xi=\eta/10$, and OBCs.}
    \label{FA7}
\end{figure}

Here we plot the fidelity  of the biorthogonal eigenpair $|\langle\Psi_0^L|\Psi_0^R\rangle|^2$ for both the 1D and the 2D $PT$-symmetric TFIM [see Fig.~\ref{FA7}]. We see that the biorthogonal eigenpair fidelity remains high with increasing system size.

\section{Energy variance}\label{A3}

\subsection{Non-Hermitian variance derivation}

\begin{figure*}[ht!]
    \centering
    \subfloat[\centering ]{{\includegraphics[width=0.49\textwidth]{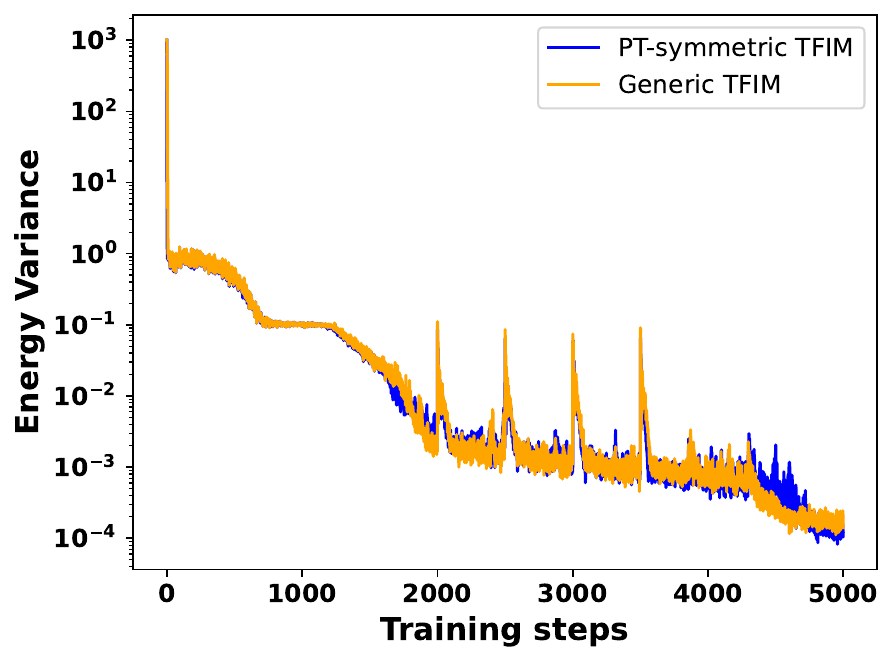} }}
    \subfloat[\centering ]{{\includegraphics[width=0.49\textwidth]{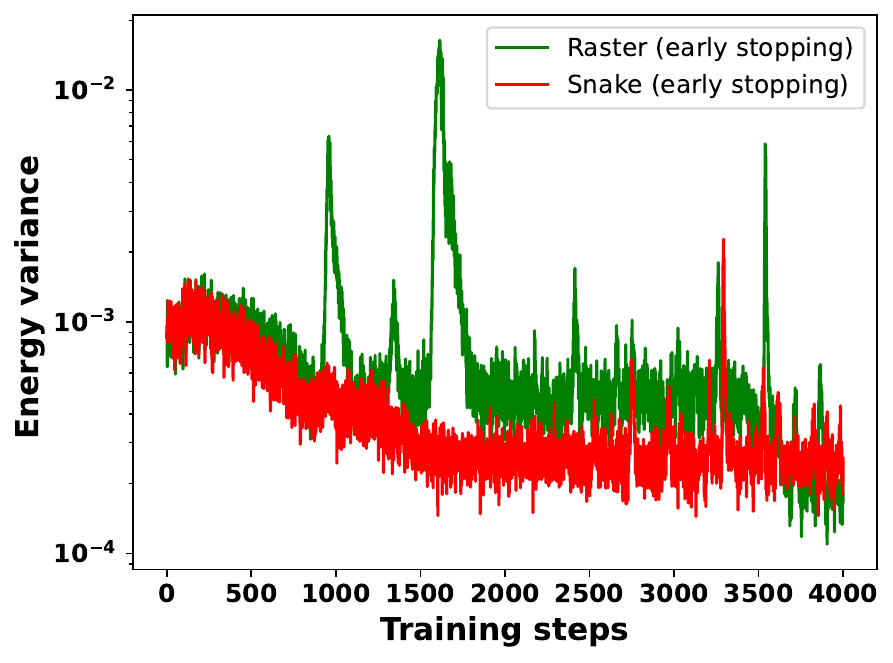} }}
    \caption{\textbf{Variance}. We plot the energy variance during training for the \textbf{(a)} 1D $PT$-symmetric TFIM (blue), and the generic TFIM (orange). We also show the energy variance for the \textbf{(b)} 2D $PT$-symmetric TFIM with early stopping for different raster (green) and snake (blue) site scanning. We consider \textbf{(a)} $N=10$, $\eta=1.6$, $\xi=\eta/10$, and \textbf{(b)} $L_x \times L_y=3 \times 3$, $\eta=1.6$, $\xi=\eta/10$. In all cases, the variance remains relatively low.}
    \label{FA3}
\end{figure*}

In standard Hermitian VMC, the energy variance serves as a natural convergence diagnostic and loss function, since it vanishes when the ansatz corresponds a true eigenstate of the Hamiltonian. In the NH context, however, this quantity is  generally complex valued (as the energy itself is generally complex) and therefore unbounded below, rendering the optimization landscape nontrivial. Instead, one can define, for instance, a right variance such that
\begin{align*}
    \bar{\sigma}_R = \frac{\langle\Psi^R|(H^{\dagger}-\varepsilon^*)(H-\varepsilon)|\Psi^R\rangle}{\langle\Psi^R|\Psi^R\rangle} = \frac{||(H-\varepsilon)|\Psi^R\rangle||^2}{\langle\Psi^R|\Psi^R\rangle} \geq 0.
\end{align*}
This is always real and nonnegative (can then serve as a loss function) since it is a square Euclidean norm ($||.||$), regardless of the Hermiticity of $H$. Expanding in the computational basis $\{\bm{x}\}$ yields
\begin{align*}
    \bar{\sigma}_R =\mathbb{E}_{\sim|\Psi^R|^2}[|E_{\text{loc}}(\bm{x})-\varepsilon|^2],
\end{align*}
where $\mathbb{E}_{\sim|\Psi^R|^2}$ is the expectation values over $\Psi^R$, and $E_{\text{loc}}(\bm{x})$ is the standard local energy \cite{wah2025} defined as
\begin{align}\label{EqA3}
    E_\textrm{loc}(\bm{x})&= \frac{\langle\bm{x}|H|\Psi_{0,R}\rangle}{\langle \bm{x}| \Psi_{0,R}\rangle}=\sum_{\bm{x}'}H_{\bm{x}\bm{x}'} \frac{\Psi^R(\bm{x}')}{\Psi^R(\bm{x})},
\end{align}
where, $\bm{x}'$ denotes the configuration with the $\text{i}^\text{th}$ spin flipped.

Interestingly, expanding $|E_{\text{loc}}(\bm{x})-\varepsilon|^2= [\text{Re}(E_{\text{loc}}(\bm{x}))-\text{Re}(\varepsilon))]^2 + [\text{Im}(E_{\text{loc}}(\bm{x}))-\text{Im}(\varepsilon))]^2$, shows that the variance receives contributions from both the real and imaginary fluctuations of the local energy. This constitutes a distinctive feature of the NH setting: even in the $PT$-unbroken phase, where the true eigenvalues are real, an inaccurate ansatz can produce spurious imaginary fluctuations in the local energy, which are automatically penalized by the variance. The left energy variance can be defined analogously. 

\subsection{Implementation}
We plot the energy variance as a function of the training steps. As shown in Fig.~\ref{FA3}, both models achieve very low variance. For the 2D TFIM [see Fig.~\ref{FA3}(b)] the snake scan exhibits smaller fluctuations, remains more stable during training, and converges faster than the raster scan.

\subsection{Biorthogonal variance}

Note that, in principle, one could also construct the variance using the biorthogonal product (as opposed to enforcing biorthogonality by summing the left and right variances). In this case, the variance would generally be complex, and its real and imaginary parts could be analyzed separately. 

In other words, we define
\begin{align*}
    \bar{\sigma}^{\text{bi}}= \frac{\langle\Psi^L|(H^{\dagger}-\varepsilon^*)(H-\varepsilon)|\Psi^R\rangle}{\langle\Psi^L|\Psi^R\rangle} = \langle H^2\rangle_{\text{bi}} -\langle H\rangle_{\text{bi}}^2 \in \mathbb{C},
\end{align*}
where $\varepsilon=\langle H\rangle_{\text{bi}}$ is updated self-consistently. In this case, we evaluate
\begin{align*}
    \bar{\sigma}^{\text{bi}} =\mathbb{E}_{\text{bi}}[|E_{\text{loc}}(\bm{x})-\varepsilon|^2],
\end{align*}
where the expectation is biorthogonaly weighted. The local energy fluctuation becomes
\begin{align*}
    E_{\text{loc}}(\bm{x})-\varepsilon &= [\text{Re}(E_{\text{loc}}(\bm{x}))-\text{Re}(\varepsilon))] \\
    &+ i[\text{Im}(E_{\text{loc}}(\bm{x}))-\text{Im}(\varepsilon))] \\
    &= \delta E^{\text{real}}(\bm{x}) +i\delta E^{\text{imag}}(\bm{x}).
\end{align*}
Squaring the above complex number yields
\begin{align*}
    \text{Re}(\bar{\sigma}^{\text{bi}})=& \mathbb{E}_{\text{bi}}[(\delta E^{\text{real}})^2-(\delta E^{\text{imag}})^2] \\ \notag
    \text{Im}(\bar{\sigma}^{\text{bi}})=& 2\mathbb{E}_{\text{bi}}[\delta E^{\text{real}}\cdot\delta E^{\text{imag}}]
\end{align*}
In this case, the loss functions defined above remain valid. The real part of the variance measures the competition between the real and imaginary parts of the local energy. In particular, it can signal the unbroken-to-broken phase transition, since it changes sign at the transition where the imaginary part of the energy dominates, and vice versa. The imaginary part, on the other hand, measures the biorthogonal covariance between the real and imaginary fluctuations of the local energy. In other words, it quantifies whether the deviations of the local energy from $\varepsilon$ along the real direction are correlated with its derivations along the imaginary direction. We thus anticipate that using a biorthogonal variance within this setting could provide a clear signature of the unbroken-to-broken phase transition. However, we leave such an extension to future work, as the study of this phase transition is beyond the scope of the present work.

\section{Simulation parameters} \label{A4}

All the parameters needed for simulations on 2D and 1D models can be found in Table~\ref{tabA41} and ~\ref{tabA4} respectively.

\begin{table*}[!ht]
\centering
\begin{tabular}{|c|c|c|c|} 
\hline
\textbf{NQS}&\textbf{Methods} & \textbf{Hyperparameters} & \textbf{Entries }\\ \hline
& &Optimizer &  Adam  \\
& &RNN cell &  GRU \\
& &Seed & $111$ \\
& &Input dimension & $2$ \\
& &Sampling & Exact autoregressive \\
& &Coupling J  & $1$ \\
& &Lattice geometry & Square lattice ($L_x \times L_y$) \\
& &System size & $3\times3$, $4\times4$, $5\times5$ \\
& &Boundary conditions & OBCs \\
& &Site ordering & Snake / Raster \\
& &Activation function & tanh/Softmax \\
& &Learning rate & $5\times10^{-4}$, $10^{-3}$ \\
Biorthogonal RNN &Adaptive&Number of hidden units & $32\to64\to128\to256\to512$ \\
& &Number of layers & $1$ \\
& &Hidden dimension& $2\to4\to ...\to128\to256$\\
& &Ansatz & 1D cRNN/ pRNN with site ordering \\
& &Number of samples & $300$-$600$ \\
& &Training steps & $5\times10^3$, $1\times10^4$ per model size\\
& &Convergence window $w$ & $100$\\
& &Warm up fraction& $0.7$\\
& &Convergence threshold & $10^{-5}$\\
& &Energy weight $\lambda^t$ & $1\to0\to1\to...(\text{adiabatically})$\\
& &Early stopping patience & $500$-$800$ steps\\
& &Early stopping threshold & $10^{-5}$\\
\hline
\end{tabular}
\caption{\textbf{Hyperparameters for 2D simulations}. Here, we record all the parameters used to run the adaptive approach for the 2D $PT$-symmetric TFIM on square lattices. All experiments were performed on an HPC cluster using NVIDIA Quadro RTX 6000 GPUs (24\,GB VRAM, CUDA~12.2).}
\label{tabA41}
\end{table*}

\begin{table*}[!ht]
\centering
\begin{tabular}{|c|c|c|c|} 
\hline
\textbf{NQS}&\textbf{Methods} & \textbf{Hyperparameters} & \textbf{Entries }\\ \hline
& &Optimizer &  Adam  \\
& &RNN cell &  Vanilla/GRU \\
& &Seed & $111$ \\
& &Input dimension & $2$ \\
& &Sampling & Exact autoregressive \\
& &Coupling J  & $1$ \\
& &System size & $10, 100$ \\
& &Activation function & tanh/Softmax \\
& &Learning rate & $10^{-3}$ \\
Biorthogonal RNN &Adaptive&Number of hidden units & $32$ \\
& &Number of layers & $1$ \\
& &Hidden dimension& $2\to4\to ...\to128\to256$\\
& &Ansatz & 1D pRNN \& cRNN \\
& &Number of samples & $100$ \\
& &Training steps & $5\times10^5$\\
& &Convergence window $w$ & $50$\\
& &Boundary conditions & OBCs \\
& &Warm up fraction& $0.5$\\
& &Convergence threshold & $10^{-5}$\\
& &Energy weight $\lambda^t$ & $1\to0\to1\to...(\text{adiabatically})$\\
\hline
& &Optimizer &  Adam \\
& &RNN cell &  Vanilla/GRU \\
& &Seed & $111$ \\
& &Input dimension & $2$ \\
& &Sampling & Exact autoregressive \\
& &Coupling J  & $1$ \\
& &System size & $10, 100$ \\
& &Activation function & tanh/Softmax \\
& &Learning rate & $10^{-3}$ \\
Biorthogonal RNN &Static&Number of hidden units & $32$ \\
& &Number of layers & $1$ \\
& &Hidden dimension& $256$\\
& &Ansatz & 1D pRNN \& cRNN \\
& &Number of samples & $300$ \\
& &Training steps & $5\times10^5$\\
& &Convergence window $w$ & $50$\\
& &Warm up fraction& $0.5$\\
& &Boundary conditions & OBCs \\
& &Convergence threshold & $10^{-5}$\\
& &Energy weight $\lambda^t$ & $1\to0\to1\to...(\text{adiabatically})$\\
\hline

\end{tabular}
\caption{\textbf{Hyperparameters for 1D simulations}. Here, we record all the parameters used to run the COMPASS approach for 1D models. All experiments were performed on an HPC cluster using NVIDIA Quadro RTX 6000 GPUs (24\,GB VRAM, CUDA~12.2).}
\label{tabA4}
\end{table*}

\begin{figure*}
    \centering
    \includegraphics[width=0.8\textwidth]{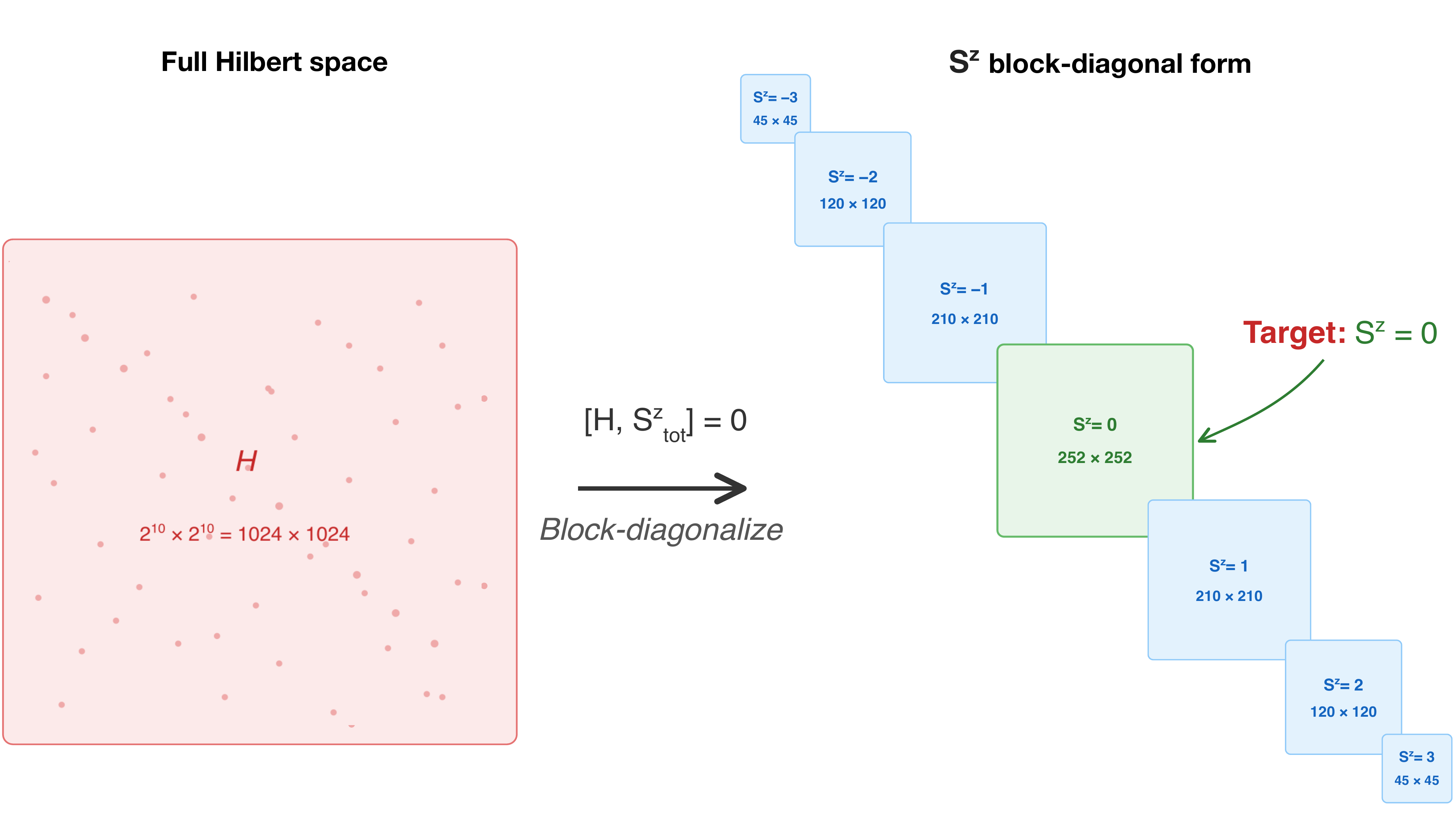}
    \caption{\textbf{Block-diagonalization via $S^z_{\text{tot}}$ conservation.} We schematically represent the block-diagonalization procedure. On the left (red box), the full Hamiltonian matrix has dimensions $2^N \times 2^N$. On the right (green and blue boxes), the block-diagonal structure is depicted with $N+1$ sectors labeled by $S^z$. The ground state resides in the $S^z=0$ sector (green), whose dimension $\begin{pmatrix} N \\ N/2 \end{pmatrix}$ is grammatically smaller than $2^N$. For illustration we consider $N=10$.}
    \label{FH}
\end{figure*}

\section{Block-diagonalization}\label{A8}

Both non-Hermitian $J_1-J_2$ models conserve $S_{\text{tot}}^z$, allowing block-diagonalization. For even $N$, the ground state lies in the $S^z=0$ sector, whose dimension $\begin{pmatrix} N \\ N/2 \end{pmatrix}$ is dramatically reduced from the full $2^N$ Hilbert space (e.g., $924$ vs. $4096$ for $N=12$; $184756$ vs. $10^6$ for $N=20$). We construct the Hamiltonian as a sparse matrix in the computational basis of fixed $S^z$ states and use full diagonalization for $N\leq 14$ and Arnoldi iteration for larger systems (still $N<20$).

In general, when a Hamiltonian $H$ commutes with an operator $\mathbb{O}$, the Hilbert space decomposes into invariant subspaces labeled by the eigenvalues of $\mathbb{O}$. Within each subspace, $H$ does not connect states belonging to different eigenvalues, so the full matrix acquires a block-diagonal structure $H = \bigoplus_\alpha H_\alpha$ where each block $H_\alpha$ acts only on the states with quantum number $\alpha$. For the $J_1-J_2$ chain, $[H, S^z_{\text{tot}}] = 0$ partitions the $2^N$-dimensional space into $N+1$ sectors labeled by $S^z = -N/2, \ldots, N/2$, with largest sector $S^z = 0$ having dimension $\binom{N}{N/2}$. Diagonalizing each block independently reduces the computational cost from $\mathcal{O}(2^{3N})$ to $\sum_\alpha \mathcal{O}(d_\alpha^3)$, which is dominated by the $S^z=0$ block [see Fig.~\ref{FH}]. Crucially, this decomposition holds for both Hermitian and NH cases since neither the staggered field, nor the complex NNN coupling mix different $S^z$ sectors.

\section{Exceptional points}\label{A9}

\begin{figure*}[ht!]
    \centering
    \subfloat[\centering ]{{\includegraphics[width=0.33\textwidth]{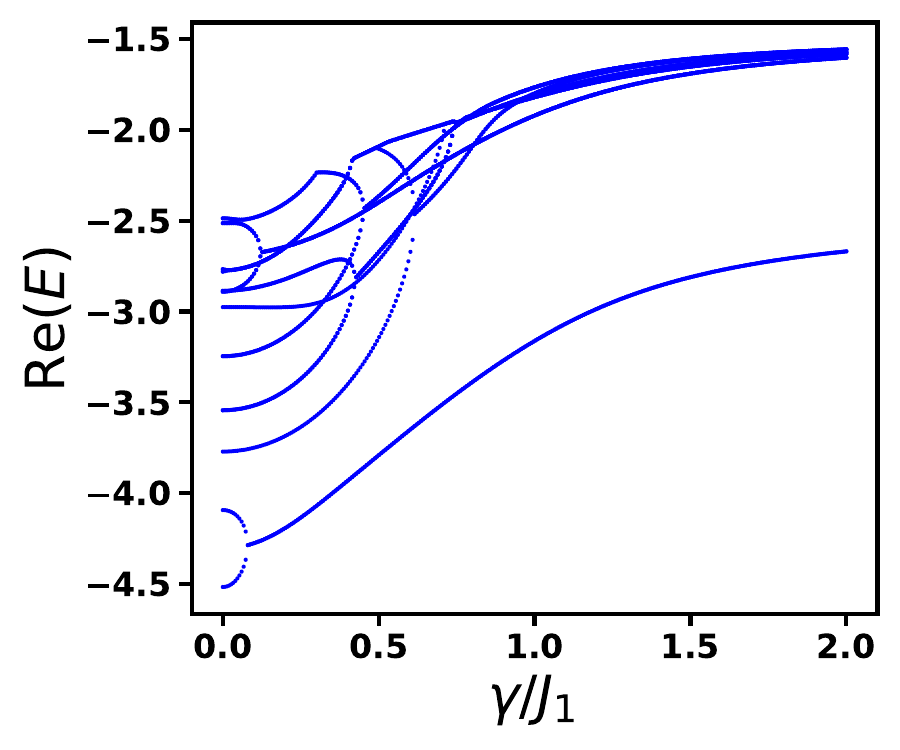} }}
    \subfloat[\centering ]{{\includegraphics[width=0.33\textwidth]{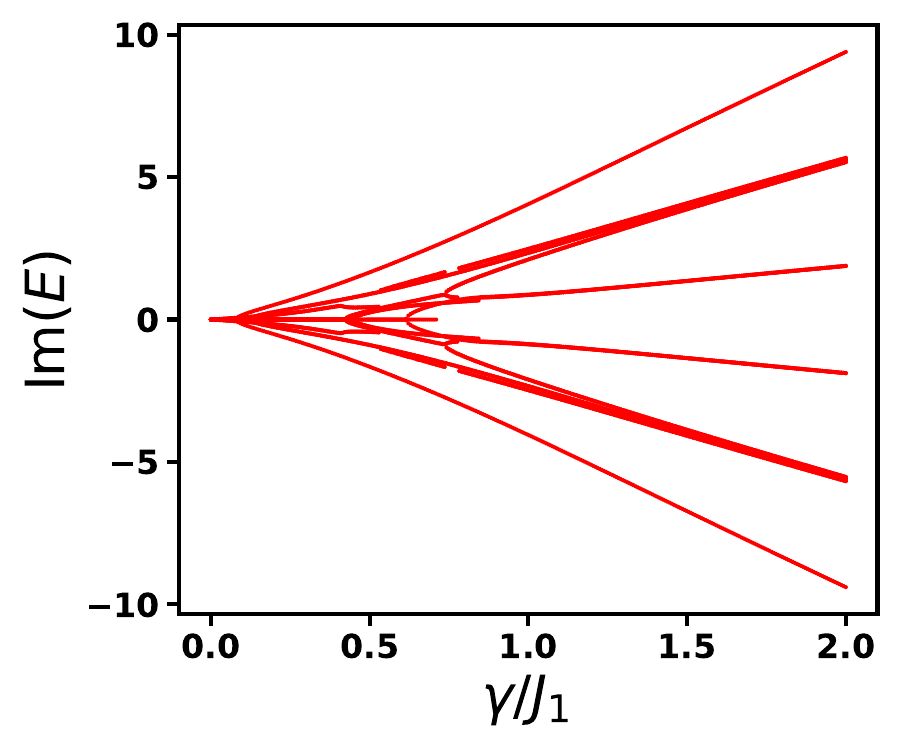} }}
    \subfloat[\centering ]{{\includegraphics[width=0.33\textwidth]{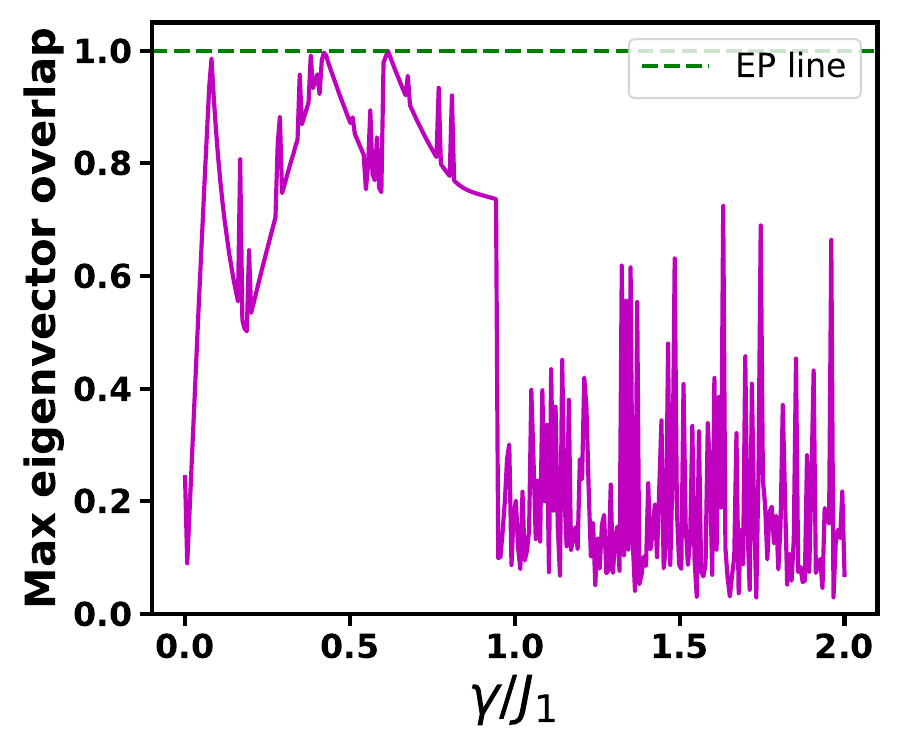} }} 

    \qquad
    \subfloat[\centering ]{{\includegraphics[width=0.33\textwidth]{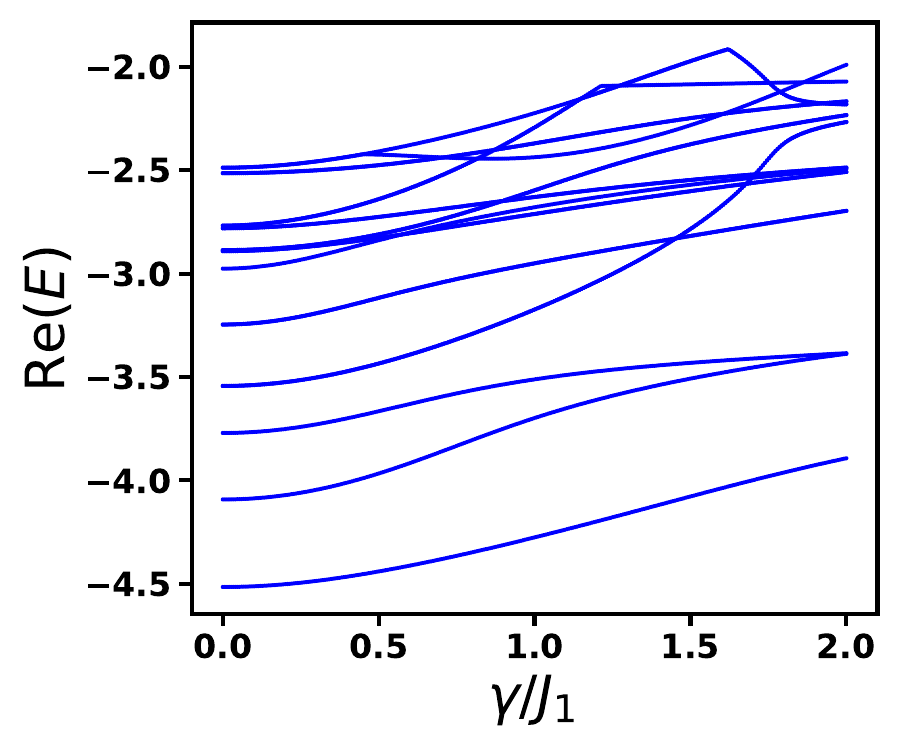} }}
    \subfloat[\centering ]{{\includegraphics[width=0.33\textwidth]{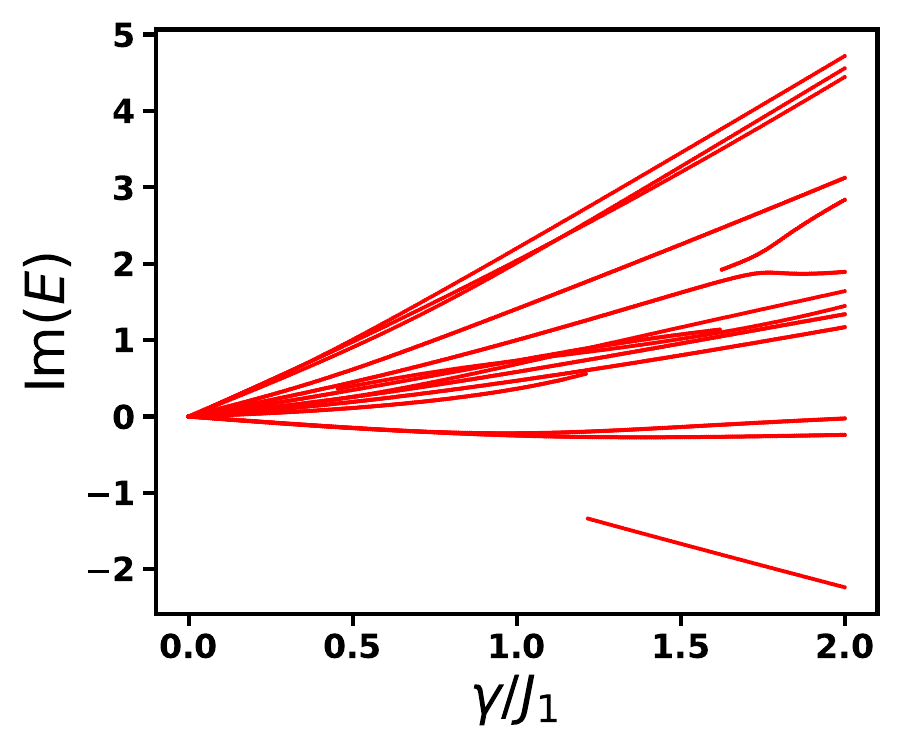} }}
    \subfloat[\centering ]{{\includegraphics[width=0.33\textwidth]{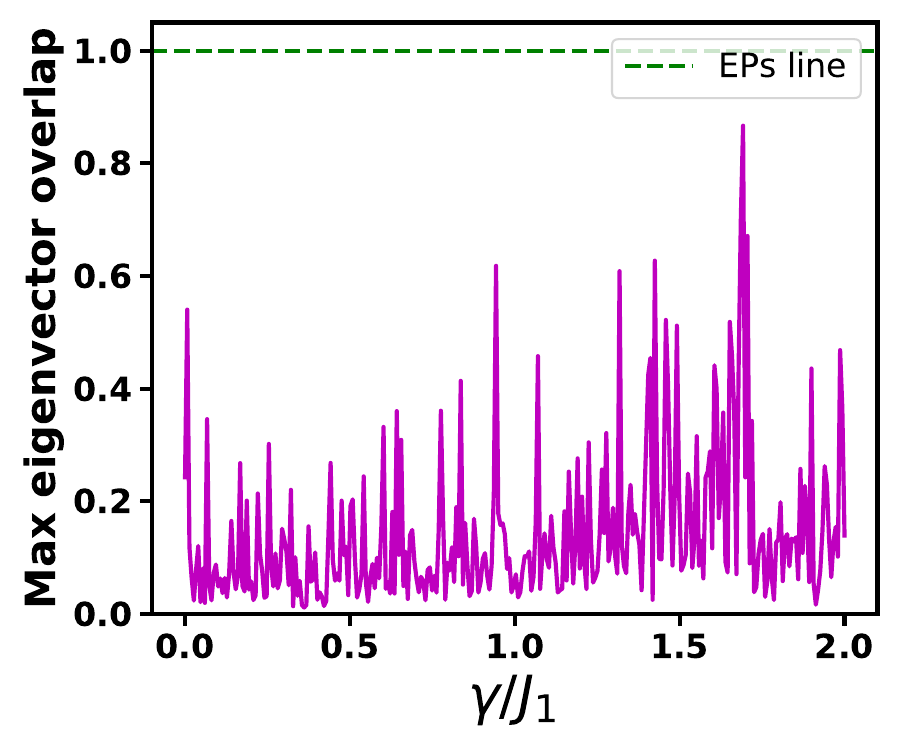} }} 
    \caption{\textbf{Emergence of exceptional points}. We plot the \textbf{(a, d)} real parts, the \textbf{(b, e)} imaginary parts of the $20$ lowest eigenenergies of \textbf{(a-c)} model $1$, and \textbf{(d-f)} model $2$. We also present the \textbf{(c, f)} maximum eigenvector overlap between all the eigenstates for both models. One observes the emergence of an EP at the same value of $\gamma/J_1$, where the maximum over lap is $1$ (green dashed line). We consider $N=10$, $J_2/J_1=0.0$ and PBCs.}
    \label{FI1}
\end{figure*}

\begin{figure*}[ht!]
    \centering
     \subfloat[\centering ]{{\includegraphics[width=0.33\textwidth]{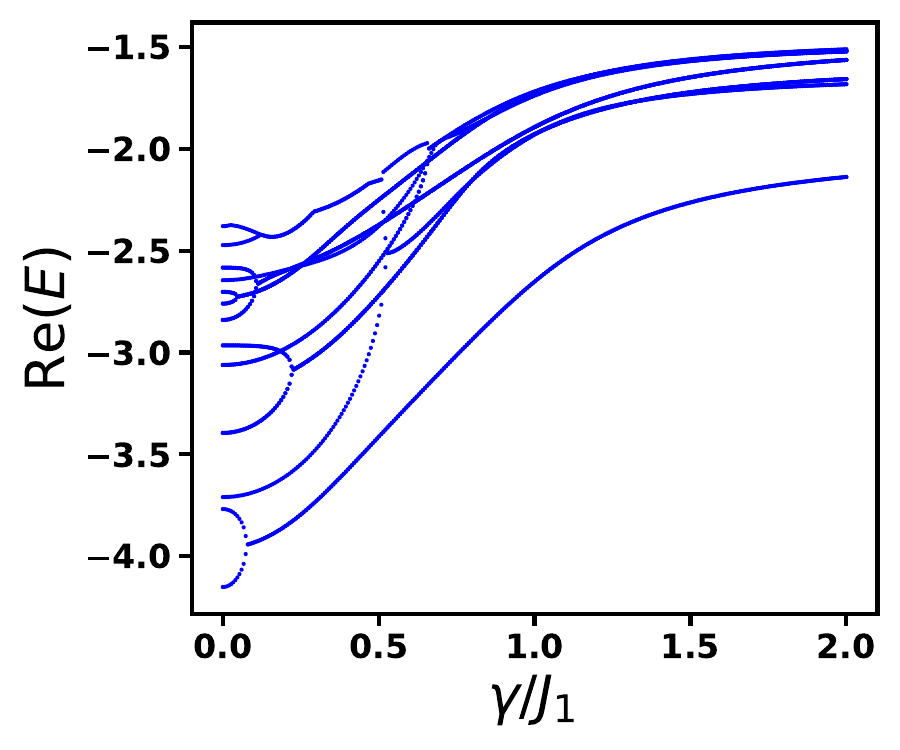} }}
    \subfloat[\centering ]{{\includegraphics[width=0.33\textwidth]{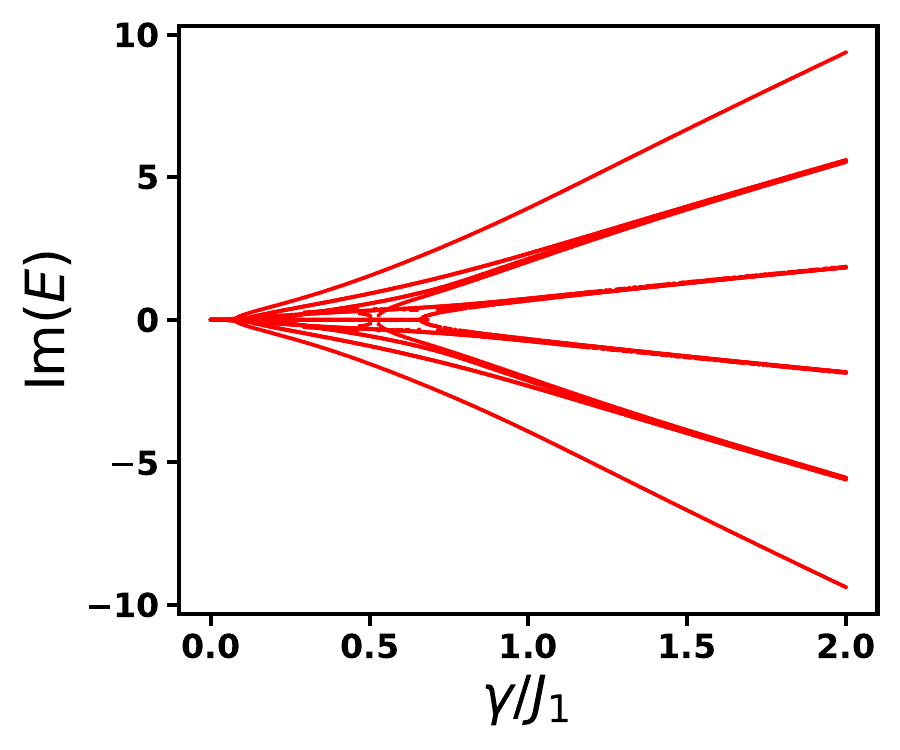} }}
    \subfloat[\centering ]{{\includegraphics[width=0.33\textwidth]{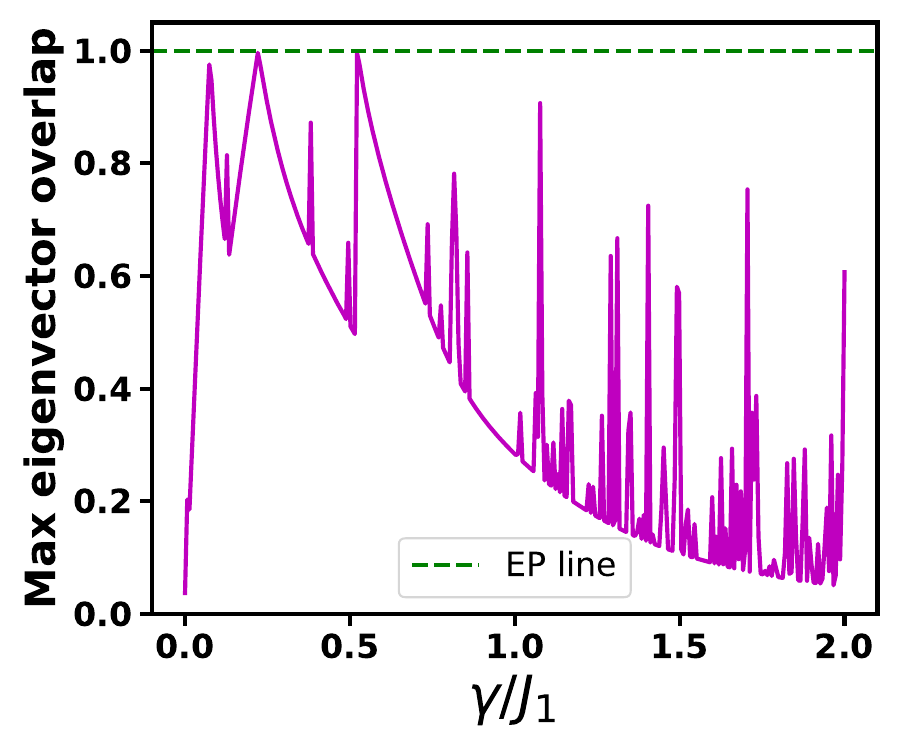} }} 

    \qquad
    \subfloat[\centering ]{{\includegraphics[width=0.33\textwidth]{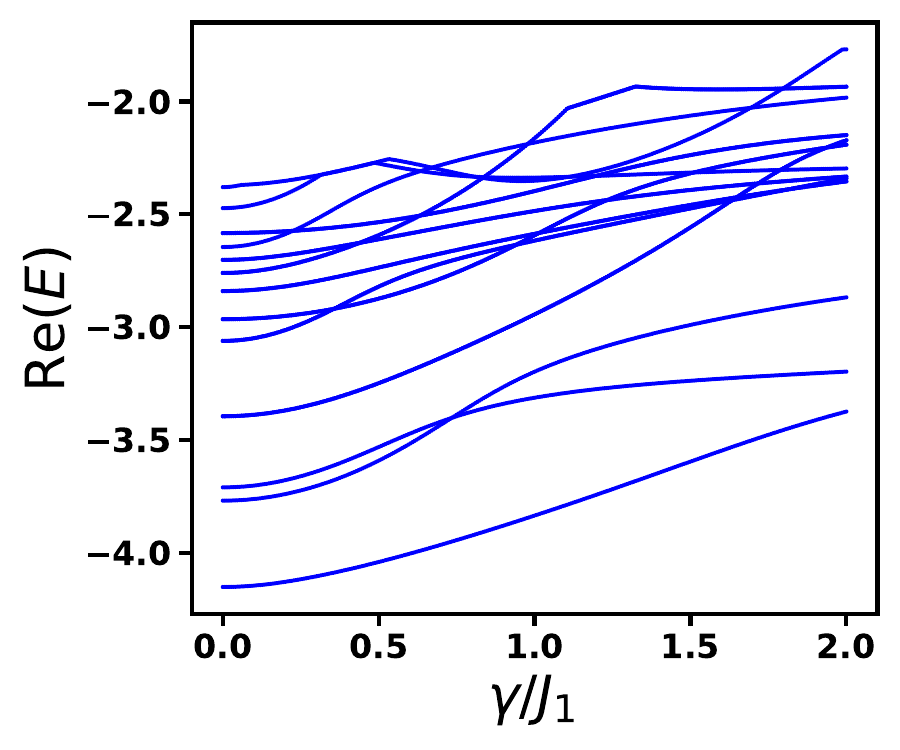} }}
    \subfloat[\centering ]{{\includegraphics[width=0.33\textwidth]{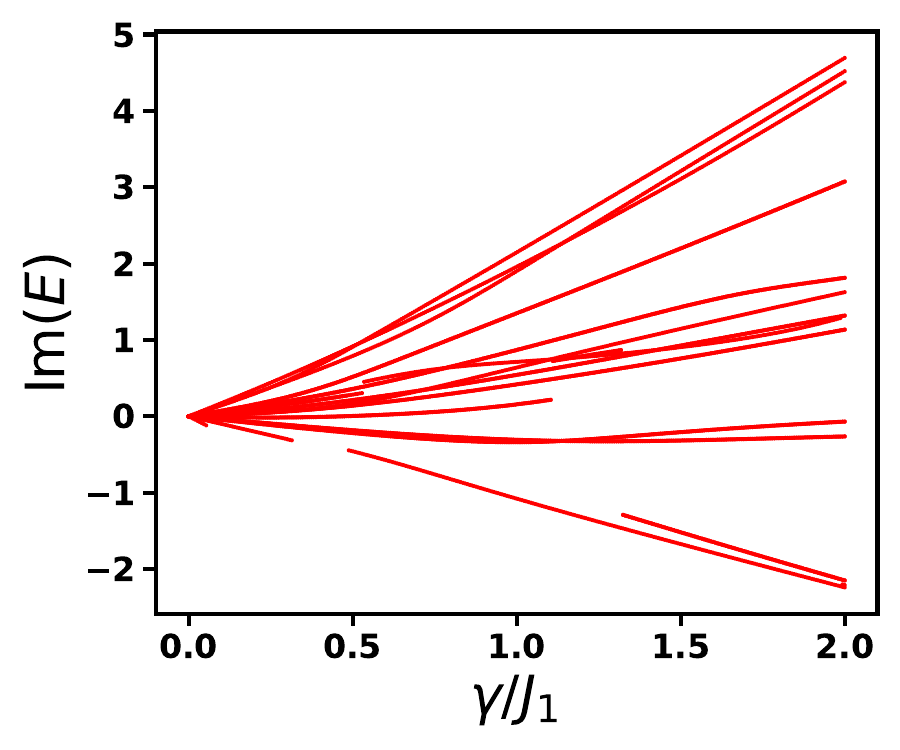} }}
    \subfloat[\centering ]{{\includegraphics[width=0.33\textwidth]{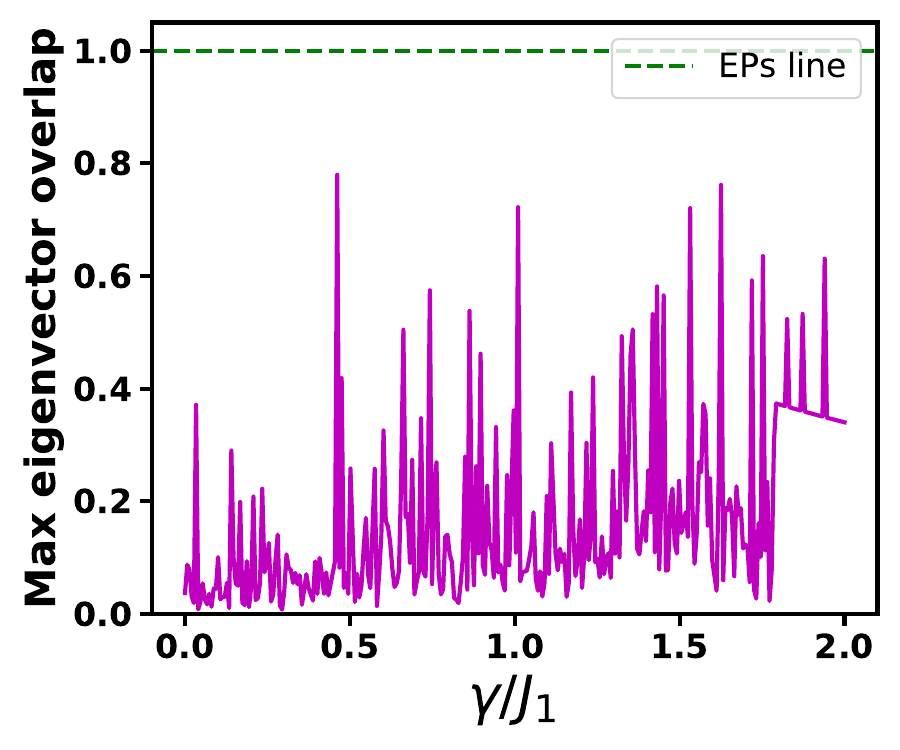} }} 
    \caption{\textbf{Emergence of exceptional points}. We plot the \textbf{(a, d)} real parts, the \textbf{(b, e)} imaginary parts of the $20$ lowest eigenenergies of \textbf{(a-c)} model $1$, and \textbf{(d-f)} model $2$. We also present the \textbf{(c, f)} maximum eigenvector overlap between all the eigenstates for both models. One observes the emergence of an EP at the same value of $\gamma/J_1$, where the maximum over lap is $1$ (green dashed line). We consider $N=10$, $J_2/J_1=0.2$ and PBCs.}
    \label{FI2}
\end{figure*}

\begin{figure*}[ht!]
    \centering
    \subfloat[\centering ]{{\includegraphics[width=0.33\textwidth]{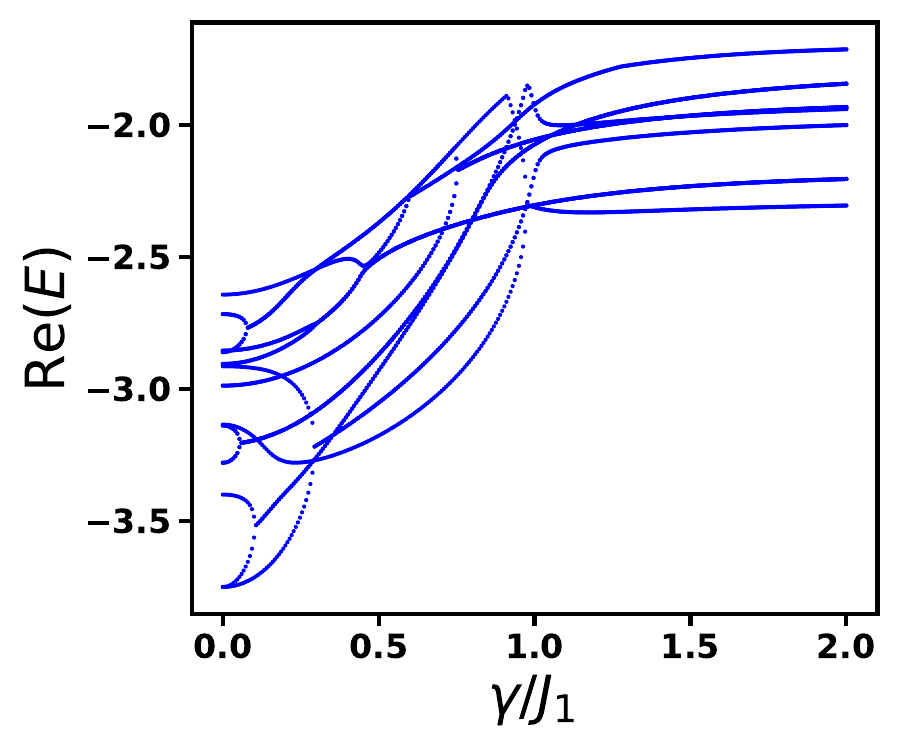} }}
    \subfloat[\centering ]{{\includegraphics[width=0.33\textwidth]{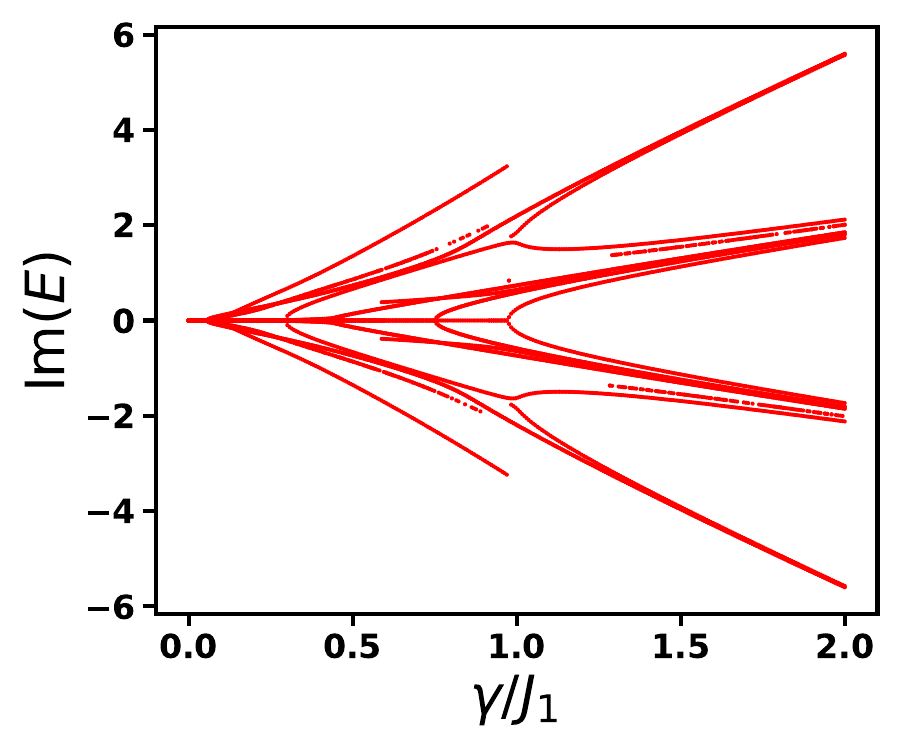} }}
    \subfloat[\centering ]{{\includegraphics[width=0.33\textwidth]{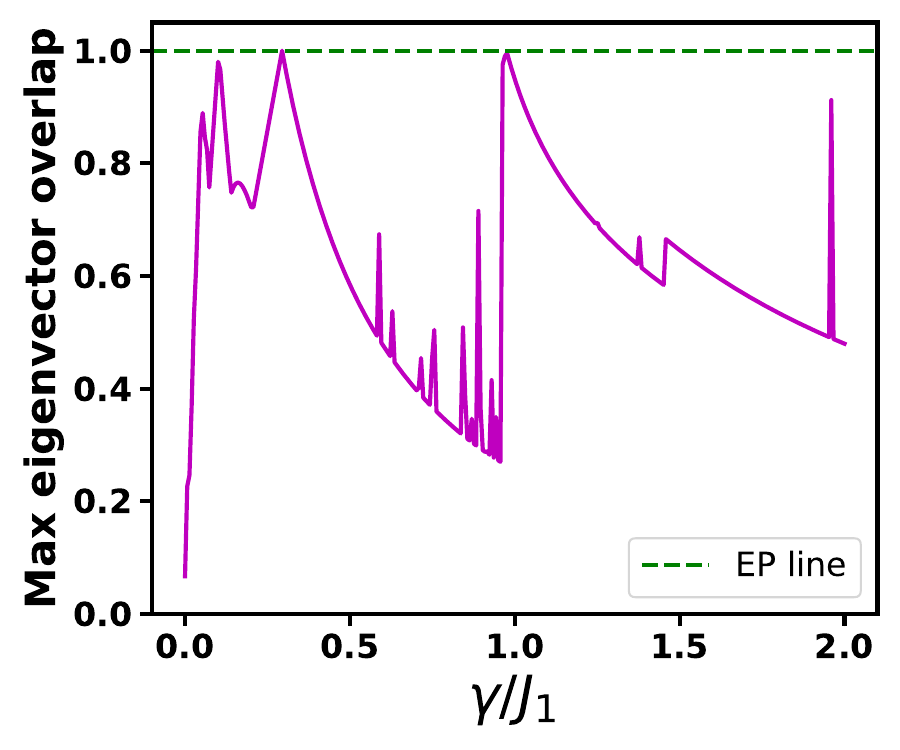} }} 
    
    \qquad
    \subfloat[\centering ]{{\includegraphics[width=0.33\textwidth]{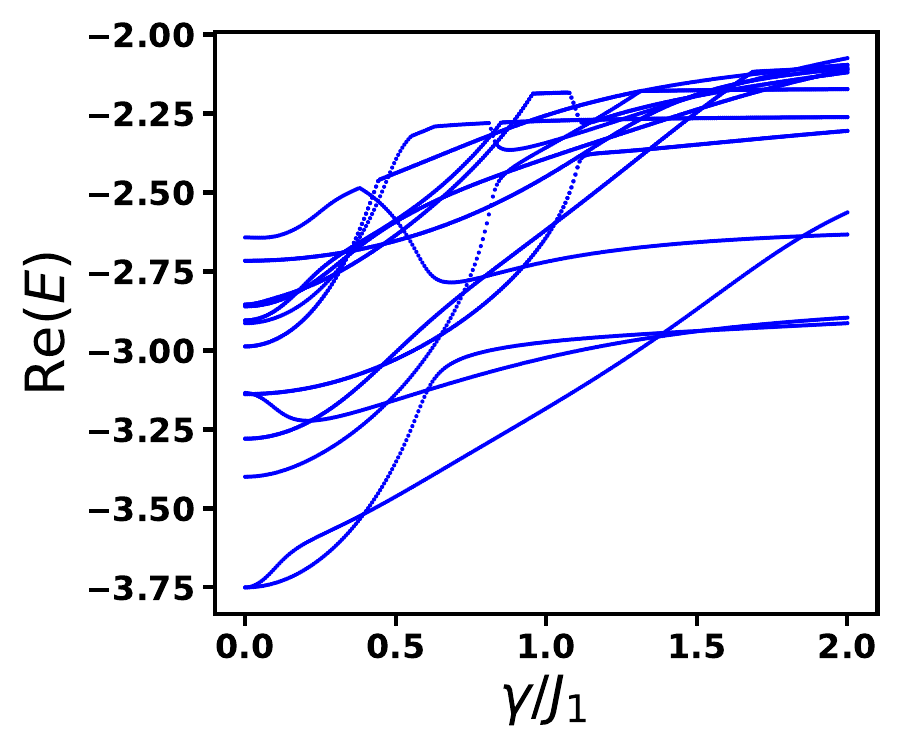} }}
    \subfloat[\centering ]{{\includegraphics[width=0.33\textwidth]{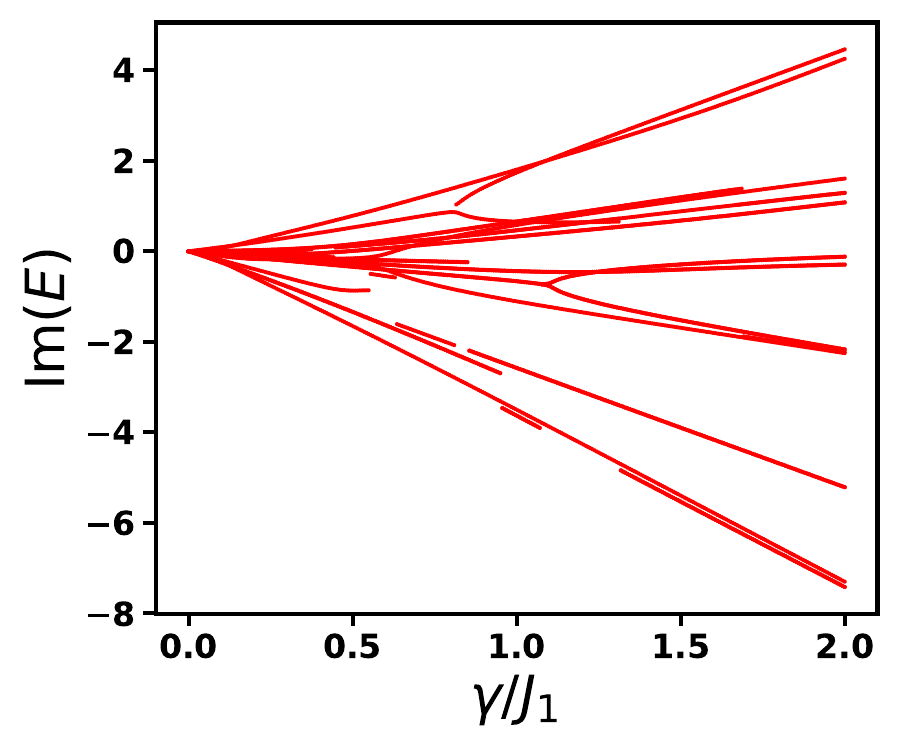} }}
    \subfloat[\centering ]{{\includegraphics[width=0.33\textwidth]{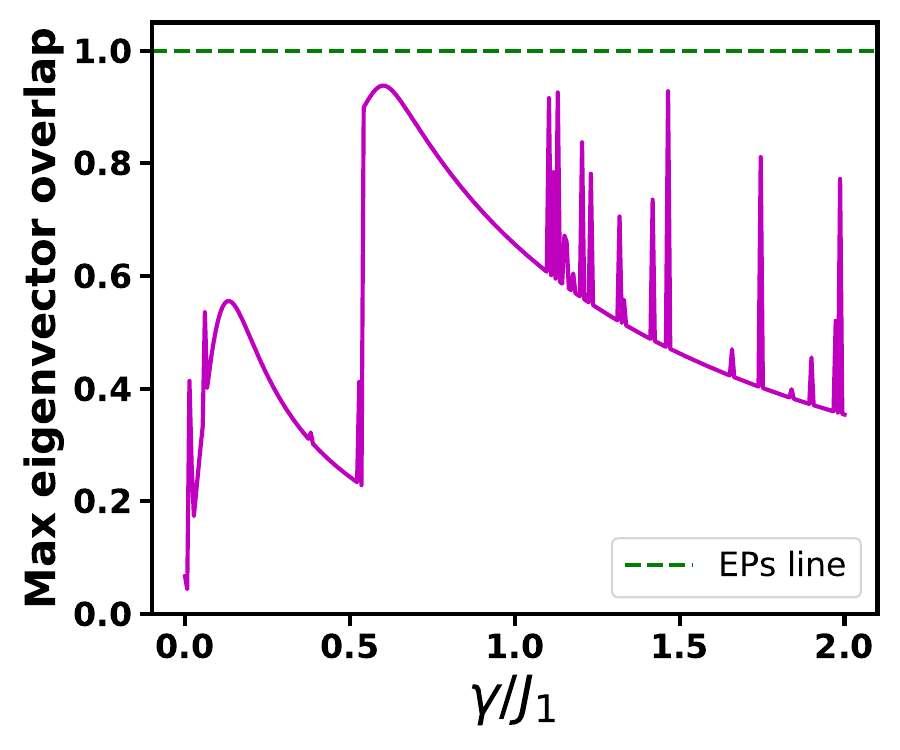} }} 
    \caption{\textbf{Emergence of exceptional points}. We plot the \textbf{(a, d)} real parts, the \textbf{(b, e)} imaginary parts of the $20$ lowest eigenenergies of \textbf{(a-c)} model $1$, and \textbf{(d-f)} model $2$. We also present the \textbf{(c, f)} maximum eigenvector overlap between all the eigenstates for both models. One observes the emergence of an EP at the same value of $\gamma/J_1$, where the maximum over lap is $1$ (green dashed line). We consider $N=10$, $J_2/J_1=0.5$ and PBCs.}
    \label{FI3}
\end{figure*}

Here we plot the real and imaginary parts of the $20$ lowest eigenenergies as a function of the non-Hermitian parameter as well as the all-to-all maximum overlap between all the eigenstates~\cite{wah2025} for both $J_1-J_2$ models at $J_2/J_1=0$ (depicted in Fig.~\ref{FI1}), $J_2/J_1=0.2$ (depicted in Fig.~\ref{FI2}), and $J_2/J_1=0.5$ (depicted in Fig.~\ref{FI3}). The emergence of exceptional points (EPs) is only observable in model $1$.

\FloatBarrier
\bibliography{bibliography}
\end{document}